\documentclass[pra,twocolumn,showpacs,superscriptaddress,floatfix, nofootinbib, showkeys]{revtex4-2}
\usepackage[caption=false]{subfig}
\usepackage{amsmath,graphicx}
\usepackage{amsmath,amssymb,mathrsfs,esint}
\usepackage{multirow}
\usepackage{algorithm}
\usepackage[noend]{algpseudocode}
\usepackage{comment}
\usepackage{tabularx}
\usepackage{bm}
\usepackage[normalem]{ulem}
\usepackage[bookmarks=true,
   colorlinks=true,
   linkcolor=blue,
   urlcolor=blue,
   citecolor=blue,
   bookmarks=true,
   hyperindex=true
]{hyperref}

\newcommand{\ket}[1]{|#1\rangle}
\newcommand{\bra}[1]{\langle#1|}

\usepackage{tikz}
\def\rx#1{\gate[][1cm]{R_x(#1)}}
\def\ry#1{\gate[][1cm]{R_y(#1)}}
\def\rz#1{\gate[][1cm]{R_z(#1)}}
\renewcommand{\r}{\ensuremath{{\sf R}}}
\renewcommand{\r}[3]{R_{#1}^{#2}(#3)}

\begin{document}

\title{Towards multiqudit quantum processor based on a $^{171}$Yb$^{+}$ ion string: \\ Realizing basic quantum algorithms}

\author{Ilia V. Zalivako}
\affiliation{P.N. Lebedev Physical Institute of the Russian Academy of Sciences, Moscow 119991, Russia}
\affiliation{Russian Quantum Center, Skolkovo, Moscow 121205, Russia}

\author{Anastasiia S. Nikolaeva}
\affiliation{Russian Quantum Center, Skolkovo, Moscow 121205, Russia}
\affiliation{P.N. Lebedev Physical Institute of the Russian Academy of Sciences, Moscow 119991, Russia}

\author{Alexander S. Borisenko}
\affiliation{P.N. Lebedev Physical Institute of the Russian Academy of Sciences, Moscow 119991, Russia}
\affiliation{Russian Quantum Center, Skolkovo, Moscow 121205, Russia}

\author{Andrei E. Korolkov}
\affiliation{P.N. Lebedev Physical Institute of the Russian Academy of Sciences, Moscow 119991, Russia}
\affiliation{Russian Quantum Center, Skolkovo, Moscow 121205, Russia}

\author{Pavel L. Sidorov}
\affiliation{P.N. Lebedev Physical Institute of the Russian Academy of Sciences, Moscow 119991, Russia}
\affiliation{Russian Quantum Center, Skolkovo, Moscow 121205, Russia}

\author{Kristina P. Galstyan}
\affiliation{P.N. Lebedev Physical Institute of the Russian Academy of Sciences, Moscow 119991, Russia}
\affiliation{Russian Quantum Center, Skolkovo, Moscow 121205, Russia}

\author{Nikita V. Semenin}
\affiliation{P.N. Lebedev Physical Institute of the Russian Academy of Sciences, Moscow 119991, Russia}
\affiliation{Russian Quantum Center, Skolkovo, Moscow 121205, Russia}

\author{Vasilii N. Smirnov}
\affiliation{P.N. Lebedev Physical Institute of the Russian Academy of Sciences, Moscow 119991, Russia}
\affiliation{Russian Quantum Center, Skolkovo, Moscow 121205, Russia}

\author{Mikhail A.~Aksenov}
\affiliation{Russian Quantum Center, Skolkovo, Moscow 121205, Russia}

\author{Konstantin M.~Makushin}
\affiliation{Russian Quantum Center, Skolkovo, Moscow 121205, Russia}

\author{Evgeniy O.~Kiktenko}
\affiliation{Russian Quantum Center, Skolkovo, Moscow 121205, Russia} 

\author{Aleksey K.~Fedorov}
\affiliation{P.N. Lebedev Physical Institute of the Russian Academy of Sciences, Moscow 119991, Russia}
\affiliation{Russian Quantum Center, Skolkovo, Moscow 121205, Russia}

\author{Ilya A. Semerikov}
\affiliation{P.N. Lebedev Physical Institute of the Russian Academy of Sciences, Moscow 119991, Russia}
\affiliation{Russian Quantum Center, Skolkovo, Moscow 121205, Russia}

\author{Ksenia Yu.~Khabarova}
\affiliation{P.N. Lebedev Physical Institute of the Russian Academy of Sciences, Moscow 119991, Russia}
\affiliation{Russian Quantum Center, Skolkovo, Moscow 121205, Russia}

\author{Nikolay N.~Kolachevsky}
\affiliation{P.N. Lebedev Physical Institute of the Russian Academy of Sciences, Moscow 119991, Russia}
\affiliation{Russian Quantum Center, Skolkovo, Moscow 121205, Russia}

\begin{abstract}
We demonstrate a quantum processor based on a 3D linear Paul trap that uses $^{171}$Yb$^{+}$ ions with 8 individually controllable four-level qudits (ququarts), which is computationally equivalent to a 16-qubit quantum processor. 
The design of the developed ion trap provides high secular frequencies, low heating rate, which, together with individual addressing and readout optical systems, allows executing quantum algorithms. 
In each of the 8 ions, we use four electronic levels coupled by E2 optical transition at 435\,nm for qudit encoding. 
We present the results of single- and two-qubit operations benchmarking
and realizing basic quantum algorithms, including Bernstein-Vazirani  and Grover's search algorithms as well as H$_2$ and LiH molecular simulations.
Our results pave the way to scalable qudit-based quantum processors using trapped ions. 
\end{abstract}


\maketitle 

\section{Introduction}

Since the conceptualisation of the trapped-ion-based quantum computing~\cite{CiracZoller1995} and first experimental demonstrations of two-qubit gates~\cite{Wineland1995,Blatt2003,Blatt2003-2}, 
this physical platform is among the leading candidates for building scalable quantum computing devices~\cite{Fedorov2022}. 
Various quantum algorithms have been tested with trapped ions,
such as quantum simulation~\cite{Blatt2012}, Deutsch--Jozsa algorithm~\cite{Blatt2003}, basic prime factorization~\cite{Blatt2016}, as well as Bernstein-Vazirani and Hidden Shift algorithms~\cite{Monroe2019}.
Trapped-ion-based quantum devices have been used to demonstrate error correction~\cite{Wineland2004,Blatt2011,Blatt2020,Monroe2021,Blatt2021-2,Postler2022}, 
e.g., a fault-tolerant entanglement between two logical qubits~\cite{Postler2022,Ryan-Anderson2022, da2024demonstration} and the quantum algorithms with logical qubits~\cite{Yamamoto2023} have been realized. 
Today trapped-ion systems show the highest quantum volume (QV) of {2097152 ($2^{21}$) in experiments by Quantinuum with the 56-qubit H2 processor~\cite{Quantinuum2022}}.
We note that superconducting circuits~\cite{Martinis2019,Pan2021-4}, semiconductor quantum dots~\cite{Vandersypen2022,Morello2022,Tarucha2022}, photonic systems (both in free space and integrated optics)~\cite{Pan2020,Lavoie2022}, 
neutral atoms~\cite{Lukin2021,Browaeys2021,Browaeys2020-2,Saffman2022} are also under development as physical platforms for quantum computing.  

However, the scaling to large enough numbers of ion qubits without the decrease of the gate fidelities remains a significant problem~\cite{Monroe2013,Sage2019}. 
The ``heart'' of an ion quantum processor is a trap, which allows one to create and control ion strings. 
A well-developed approach is to use a linear Paul trap~\cite{Monroe2013}. 
Three-dimensional volumetric trap design with metal blades or rods is widely used in research laboratories for 
manipulation with several tens of ion qubits providing long coherence times~\cite{Kim2021}, high fidelity of quantum gates~\cite{Wineland2016}, and all-to-all connectivity~\cite{Monroe2013,Brown2021}. 
However, scaling to hundreds of ions within this approach remains challenging. 
Specifically, increasing the number of ions saturates the spectrum of vibrational modes, used as a quantum bus between particles, leading to the decrease of the ion-to-ion entanglement fidelity.
In order to overcome this challenge, instead of increasing the number of particles within a single trap, one can use several ion traps and share entanglement between them. 
This can be achieved by employing photonic interconnects between remote traps~\cite{krutyanskiy2023entanglement, stephenson2020high} or by so-called quantum charge-coupled device architecture~\cite{Monroe2002}, in which ions are physically transported between separated traps~\cite{Foss-Feig2021}. Both these approaches have been demonstrated experimentally, however, they appear to be quite technologically demanding~\cite{Foss-Feig2021,Brown2021}.

An inherent path to scale trapped-ion-based quantum processors is to use multilevel encoding. 
Indeed, each of the ions admits encoding not only qubits but also qudits, namely $d$-level quantum systems, which are widely studied 
in the context of quantum technologies~\cite{Farhi1998,Kessel1999,Kessel2000,Kessel2002,Muthukrishnan2000,Nielsen2002,Berry2002,Klimov2003,Bagan2003,Vlasov2003,Clark2004,Leary2006,Ralph2007,White2008,Ionicioiu2009,Ivanov2012,Li2013,Kiktenko2015,Kiktenko2015-2, Song2016,Frydryszak2017,Bocharov2017,Gokhale2019,Pan2019,Low2020,Jin2021,Martinis2009,White2009,Wallraff2012,Mischuck2012,Gustavsson2015,Martinis2014,Ustinov2015, Morandotti2017,Balestro2017,Low2020,Sawant2020,Senko2020,Pavlidis2021,Rambow2021,OBrien2022,Nikolaeva2022,Nikolaeva2023}. 
In the context of quantum computing, qudits can be used, first, to encode several qubits in a single qudit~\cite{Ustinov2015,Kiktenko2015,Kiktenko2015-2}, 
and to use additional qudit levels to substitute ancilla qubits in multiqubit gate decompositions~\cite{White2009,Wallraff2012,Nikolaeva2022,Nikolaeva2023} (see also Ref.~\cite{Barenco1995}). The latter is important, for example, in the case of the Toffoli gate implementation~\cite{Ralph2007,Ionicioiu2009,Kwek2020,Baker2020,Kiktenko2020,Kwek2021,Galda2021,Gu2021}, which was experimentally realized within the use of ancillary levels on ions \cite{nikolaeva2024iontoffoli, Blatt2009Toffoli}, superconducting circuits \cite{chu2023scalable, Wallraff2012} and photons \cite{White2009}.
These approaches can be combined~\cite{Nikolaeva2021,Kiktenko2023}, in particular, in the realization of the Grover's algorithm~\cite{Nikolaeva2023}, where the use of qudits leads to significant advantages.
{Qudits have also a great potential for quantum matter simulations \cite{Chicco2023, Tacchino2021, Vezvaee2024fh}. Specifically, the ion platform has proven itself particularly well in lattice gauge theory experimental simulations with qudits \cite{Popov2024, Calaj2024, Meth2024}.
}

Recently, multiqudit processors have been realized with the use of trapped ions~\cite{Ringbauer2021,Kolachevsky2022}, superconducting circuits~\cite{Hill2021,Schuster2022, chu2023scalable}, and optical systems~\cite{OBrien2022}.
In the case of trapped ions qudit processors with up to seven levels in each qudit with high-enough gate fidelities have been shown~\cite{Ringbauer2021,Kolachevsky2022}. Control over ion qudits with even higher dimensions was also demonstrated experimentally~\cite{Low2023}.
The crucial challenges in the development of qudit ion processors are a more complicated readout procedure and the reduced coherence, as it is usually difficult to find many controllable magnetically-insensitive states in the ion. Thus, qudits are usually more susceptible to the magnetic field noise. Several approaches, including dynamical decoupling~\cite{Zalivako2023} and magnetic shielding~\cite{Ringbauer2021} are suggested to improve the coherence time, which significantly increases the qudits applicability for realizing quantum algorithms.

In this work, we present the realization of the next generation of quantum processor based on $^{171}$Yb$^{+}$ qudits, where we encode quantum information in an E2 optical transition at 435~nm. 
One of the major improvements over our previous system~\cite{Kolachevsky2022} is a new trap. 
The developed ion trap provides better optical access, higher secular frequencies as well as its architecture allows cooling, which significantly improved radial secular frequencies stability, the heating rate, and the lifetime of the ion string. 
Additional modifications include a new readout system based on a fiber array, providing individual ion-resolution and higher detection efficiency, an improved ions addressing system and a control system.
That enabled us to control a universal 8-qudit quantum processor with four levels in each particle. {As qudits can be seen as a generalization of qubits, the processor can be operated both in a full qudit regime and in a conventional qubit mode, helping to independently and progressively benchmark the different system's features.}
In addition to the details on the processor design, we also demonstrate the results of its benchmarking.
We realize basic quantum algorithms, including the Bernstein-Vazirani algorithm, Grover's search as well as H$_2$ and LiH molecular simulations.

Our work is organized as follows. 
In Sec.~\ref{sec:ytterbium}, we discuss the chosen ion species and information encoding type.
In Sec.~\ref{sec:setup}, we describe our experimental setup. 
In Sec.~\ref{sec:gates-benchmarking}, we describe benchmarking of single- and two-qudit gates in our setup.
In Sec.~\ref{sec:algorithms}, we present the results of realization of basic quantum algorithms.  
We summarize our results in Sec.~\ref{sec:conclusions}.

\section{Qudits encoded in $^{171}$Yb$^{+}$ ions} \label{sec:ytterbium}

\begin{figure}
\center{\includegraphics[width=1\linewidth]{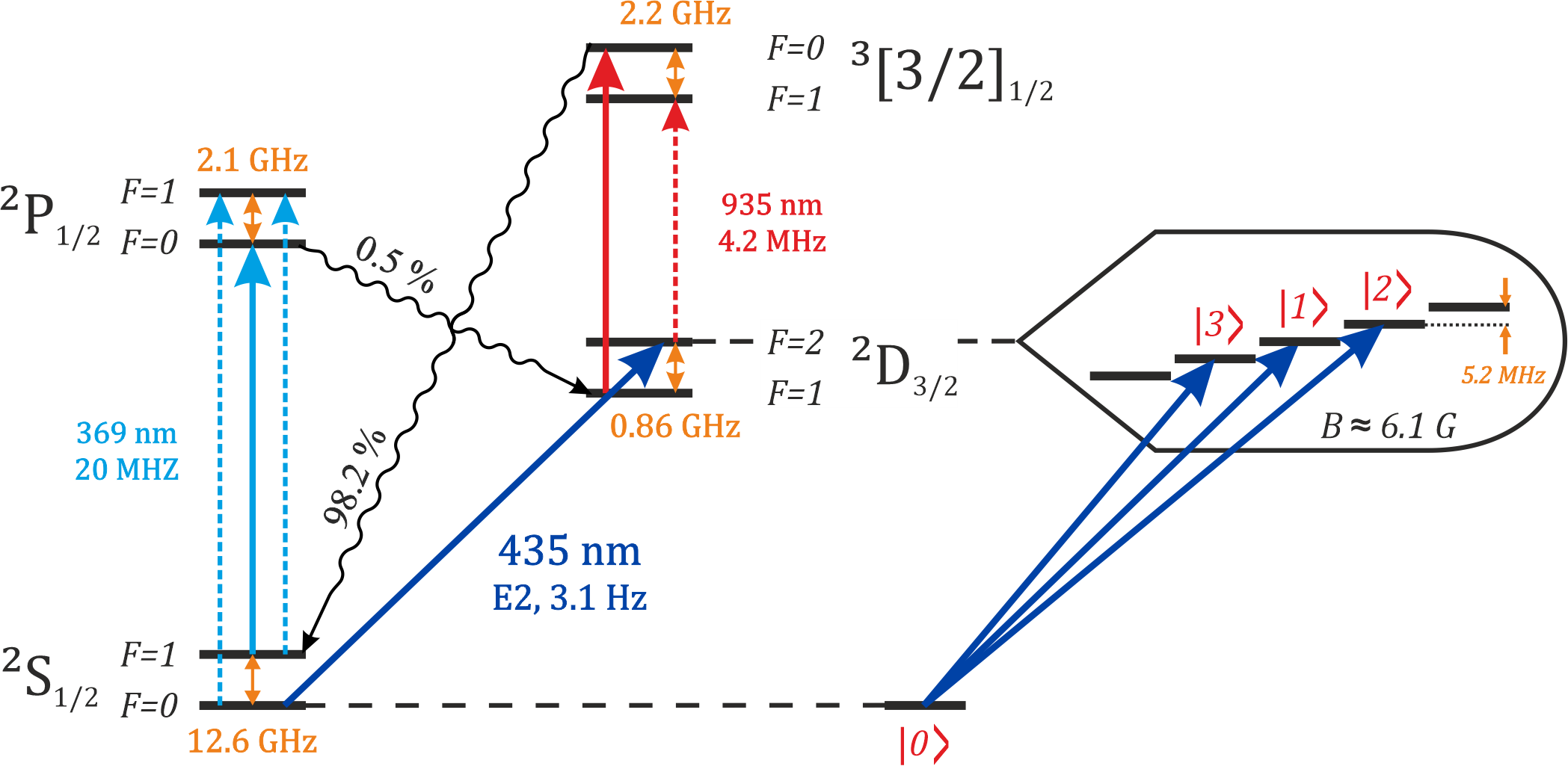}}
	\vskip-3mm
\caption{$^{171}$Yb$^{+}$ level structure. The laser fields used for laser cooling, state initialization, manipulation, and readout are shown with solid lines. Dotted lines correspond to the modulation sidebands that are obtained with electro-optical modulators and prevent population trapping in metastable hyperfine sublevels.}
\label{fig:levels}
\end{figure}

$^{171}$Yb$^{+}$ ions are one of the natural choices for the development of quantum computing devices due to their rich  and convenient energy structure~\cite{Sage2019}, see Fig.~\ref{fig:levels}. 
Laser cooling, state initialization, and high-fidelity readout can be straightforwardly implemented using readily available diode lasers~\cite{Kolachevsky2019}. 
Another feature of this type of ions is the high resistance of the ion chain to collisions with the background gas. If an Yb$^+$ ion experiences a collision with a gas particle (typically hydrogen) with the occasional formation of  YbH$^+$ molecule, the latter can be efficiently photodissociated using, e.g., cooling laser at 369\,nm. But the most important is a wide spectrum of states suitable for quantum information encoding, especially an $^{171}$Yb$^{+}$ isotope with hyperfine doublets (nuclear spin equals $I=1/2$).
Using different levels, a variety of methods both to encode quantum information and to perform quantum operations can be implemented. Today the $^{171}$Yb$^{+}$ is one of the most widely used ions for quantum processors~\cite{Monroe2013,Sage2019}, with several world-best results, e.g., the highest quantum volume~\cite{Quantinuum2023} and the longest coherence time~\cite{Kim2021}. 

In contrast to the ground state hyperfine levels encoding at 12.6\,GHz demonstrated in several experiments~\cite{Monroe2019, Moses2023}, we use an alternative approach to encode quantum information: we employ the ground state $^2S_{1/2}(F=0,m_F=0)$ and the Zeeman sublevels of $^2D_{3/2}(F=2)$ as a qudit states (Fig.~\ref{fig:levels}). Therefore, the maximum qudit dimension that is supported by our system equals $d=6$. The $|0\rangle=\,^2S_{1/2}(F=0,m_F=0)$ state can be coupled to the upper states by the electric quadrupole transition at $\lambda=435.5\,$nm with a natural linewidth of 3~Hz (the upper states lifetime equals to $\tau=53\,$ms). In this work we use only four of these states as in this way a direct mapping between our qudit system and an analogous qubit setup can be easily made. We label these states as $|0\rangle=\,^2S_{1/2}(F=0,m_F=0), |1\rangle=\,^2D_{3/2}(F=2,m_F=0), |2\rangle=\,^2D_{3/2}(F=2,m_F=1), |3\rangle=\,^2D_{3/2}(F=2,m_F=-1)$ as it is shown in Fig.~\ref{fig:levels}.
Compared to the conventional scheme with a 12.6\,GHz microwave qubit, our approach provides: (i) a larger qudit dimension; (ii) a straightforward way to readout state of the whole qudit in one shot; (iii) a convenient laser wavelength (435.5~nm) required for the qudit manipulation. 
The spectral region around 400\,nm allows combining both high spatial resolution for the individual addressing and compatibility with the visible-range optical components, including wide scan-range ${\rm TeO_2}$ acousto-optical deflectors (AODs). 
The main drawback of optical encoding is the limited qudit coherence time due to, first, the limited upper state lifetime and, second, the first-order sensitivity to the magnetic field fluctuations of the states with a non-zero magnetic quantum number. 
Still, the natural lifetime is sufficiently long to perform in principle hundreds of quantum operations, while the problem of magnetic sensitivity can be circumvented by magnetic shielding or by dynamical decoupling. 
Previously we have demonstrated experimentally coherence times of more than 9~ms for magnetic-sensitive transitions using continuous dynamical decoupling~\cite{Zalivako2023} without magnetic shielding.

\section{8-qudit quantum processor}\label{sec:setup}
\begin{figure}
\center{\includegraphics[width=1\linewidth]{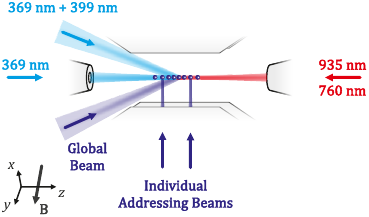}}
	\vskip-3mm
\caption{Linear Paul trap and laser beams configuration. 
Beams along the trap axis $z$ perform readout, recrystallization (369~nm) and repumping (935~nm and 760~nm). The photoionization (399~nm) and the main Doppler cooling (369~nm) beams propagate at an angle to all three principle trap axes. Individual and global addressing beams at 435~nm propagate orthogonally to the trap axis and to each other. The quantization magnetic field is directed orthogonally to the trap axis and at $60^\circ$ to the individual addressing beams.}
\label{fig:beams}
\end{figure}

{
The general setup scheme is shown in Fig.~\ref{fig:beams}. The quantum register of eight $^{171}$Yb$^{+}$ ions is stored in a four-blade linear Paul trap. The trap secular frequencies of center-of-mass motional modes along the trap principal axes are $\{\omega_x, \omega_y, \omega_z\} = 2\pi \times \{3.7, 3.8, 0.116\}$~MHz. The detailed description of the trap and all experimental procedures are given in the Methods section of the Suppl. materials. Before each experimental run ions are Doppler-cooled~\cite{Letokhov1995, Schreck2021} which is followed by the resolved-sideband cooling~\cite{Monroe1995} of radial modes to the motional ground state. The achieved mean phonon number in all radial modes is less than 0.1, while the heating rate of center-of-mass radial modes is measured to be $\dot{n}=23 \pm 3$~phonons/s (see Suppl. materials). After the cooling procedure ions are initialized in the state $\ket{0}$ using optical pumping.}

{
The quantum operations on the quantum register are performed with a laser at 435.5~nm, frequency-stabilized with respect to a high-finesse cavity. All the addressing beams are aligned orthogonally to the trap axis. One of them hits all the ions simultaneously (global addressing beam) and is used for the ground state cooling, global quantum gates and micromotion compensation (details on these procedures are given in the Suppl. materials). The other two beams (individual addressing beams --- IABs) are aligned orthogonally to the global beam and are tightly focused on the ion string to resolve individual particles. The scanning of the beams along the ion chain is performed using acousto-optical deflectors (AODs). In contrast to other works~\cite{Blatt2021-3, Ringbauer2021} we use two dedicated individual beams controlled with their own AOMs and AODs for single- and two-qudit operations instead of multi-frequency drive of an AOD in a single beam. This results in more degrees of freedom at the cost of slow relative phase fluctuations between these beams due to drifts in the non-common optical path lengths. The influence of these fluctuations is suppressed by performing operations, which could be affected, in a phase-insensitive way (see Suppl. materials). The cross-talk of the individual addressing beams are 3-5\%, depending on the ion.}

{
Two types of native single-qudit operations are supported by our system. The first type is given by the operator
\begin{equation}
    \r{\phi}{0j}{\theta}=\exp(-\imath\sigma^{0j}_{\phi}\theta/2),
\end{equation}
where $\theta$ is the rotation angle, $0$ and $j\in\{1,2,3\}$ denote addressed levels of the qudit, $\sigma_\phi^{0j} = \cos\phi \sigma_x^{0j} + \sin\phi \sigma_y^{0j}$, 
$\sigma^{0j}_\kappa$ with $\kappa = x,y,z$ stands for the standard Pauli matrix acting in the two-level subspace spanned by $\ket{0}$ and $\ket{j}$ (e.g., $\sigma_y^{03}=-\imath\ket{0}\!\bra{3}+\imath\ket{3}\!\bra{0}$), and angle $\phi$ specifies a rotation axis.
We fix notations for rotations around $x$- and $y$-axes:
$\r{x}{0j}{\theta}:= \r{\phi=0}{0j}{\theta}$, $\r{y}{0j}{\theta}:= \r{\phi=\pi/2}{0j}{\theta}$.
For rotations performed in the qubit subspace of the qudit we also fix the notation $\r{\phi}{}{\theta}:= \r{\phi}{01}{\theta}$.}

{
The second type of native single-qudit gate is a virtual phase gate:
\begin{equation}
    \r{z}{j}{\theta}=\exp\left(\imath\theta\ket{j}\bra{j}\right),
\end{equation}
which is the generalization of widely-used qubit virtual phase gates~\cite{McKay2017}. 
Here $\ket{j}=0,...,3$ is the qudit state, which acquires additional phase $\theta$. 
This gate is implemented by the appropriate shift of the IAB phase during all successive real gates and always has a fidelity of 1.}

As it is shown in Ref.~\cite{nikolaeva2023universal}, the two-qubit Hilbert space maps onto a single ququart and the universal two-qubit gate set can be constructed from ion-native single-qudit gates. 
This illustrates the fact that each qudit in our system can be treated as a pair of qubits. 

{
To entangle several qudits with each other we use the M$\o$lmer-S$\o$rensen gate~\cite{Blatt2003-2,Molmer-Sorensen1999,Molmer-Sorensen1999-2,Molmer-Sorensen2000} on a $|0\rangle\to|1\rangle$ transition. 
The gate is defined as follows:
\begin{equation}
    XX(\chi)=\exp\left(-\imath\frac{\chi}{2}(\sigma^{01}_x\otimes\mathbb{I}+\mathbb{I}\otimes\sigma^{01}_x)^2\right).
\end{equation}
In a qubit subspace, this becomes fully entangling at $\chi=\pi/4$. It should be noted that this gate does not simply reduce to the $\exp(-\imath\chi\sigma^{01}_x\otimes\sigma^{01}_x)$ but additional phases are also acquired by states $|2\rangle$ and $|3\rangle$. 
These additional phases, however, can be compensated with virtual single-qudit gates~\cite{Ringbauer2021}, which do not reduce the overall entangling gate performance. }

{
The technical details on the implementation of both single- and two-qudit operations are given in the Suppl. materials.
}

{
At the end of each experimental run, states of all qudits in the register are detected using a sequential electron-shelving technique. On the first stage, a usual electron-shelving procedure is performed to distinguish state $\ket{0}$ from others: an ion in state $\ket{0}$ strongly scatters photons under the illumination with 369~nm and 935~nm lasers, while for ions in other states the scattering is suppressed. The scattered photons are collected with an in-vacuum composite lens. An ion string image is projected by the lens on a grid of multimode optical fibers, spaced accordingly to the particles in the chain, coupled to the individual channels of a multi-channel photomultiplier tube (PMT). After the first stage, a $R_x^{01}(\pi)$ gate is applied to all the ions in the register and the measurement is repeated, while in this case the strong fluorescence appears if ion initially was in $\ket{0}$ or $\ket{1}$ state. The same is repeated again with the state $\ket{2}$. The population in the remaining state $\ket{3}$ is derived using a normalization condition. The mean state preparation and measurement (SPAM) fidelity was estimated to be 96.4(2)\% per ion. More details on the procedure and SPAM fidelity measurement data are presented in Suppl. materials.}

\section{Components benchmarking}\label{sec:gates-benchmarking}
\subsection{Single-qudit gates}

Single-qudit gates are characterized using randomized benchmarking (RB)~\cite{Wineland2008} method. 
We individually benchmark gates on each qudit transition ($|0\rangle\to|1\rangle$, $|0\rangle\to|2\rangle$, $|0\rangle\to|3\rangle$) using the classical qubit RB technique, while keeping all 8 ions in the trap. 
Cross-talk errors are ignored in this measurement.

Random circuits have been generated and then transpiled to our native gates. 
To keep the fidelities of each Clifford gate on the same level, we transpiled each gate using U3 decomposition
\begin{equation}
    \begin{aligned}
            U3(\theta, \phi, \lambda)=\begin{pmatrix}
                    \cos{\frac{\theta}{2}} & -e^{\imath\lambda}\sin{\frac{\theta}{2}}\\ 
                    e^{\imath\phi}\sin{\frac{\theta}{2}} & e^{\imath(\phi+\lambda)}\cos{\frac{\theta}{2}}
        \end{pmatrix} \to \\ \\
        R_z^{0j}(\phi)R_x^{0j}(-\pi/2)R_z^{0j}(\theta)R_x^{0j}(\pi/2)R_z^{0j}(\lambda).
    \end{aligned}
\end{equation}
Thus, each Clifford gate always has two real $\pi/2$ rotations and three virtual phase rotations. As we expect all real rotations to have approximately the same fidelity, all Cliffords, even the identity gate, have close to each other fidelities.

\begin{figure*}
\centering
\subfloat[\label{fig:SQ_RB_00}]{%
    \includegraphics[width=0.49\textwidth]{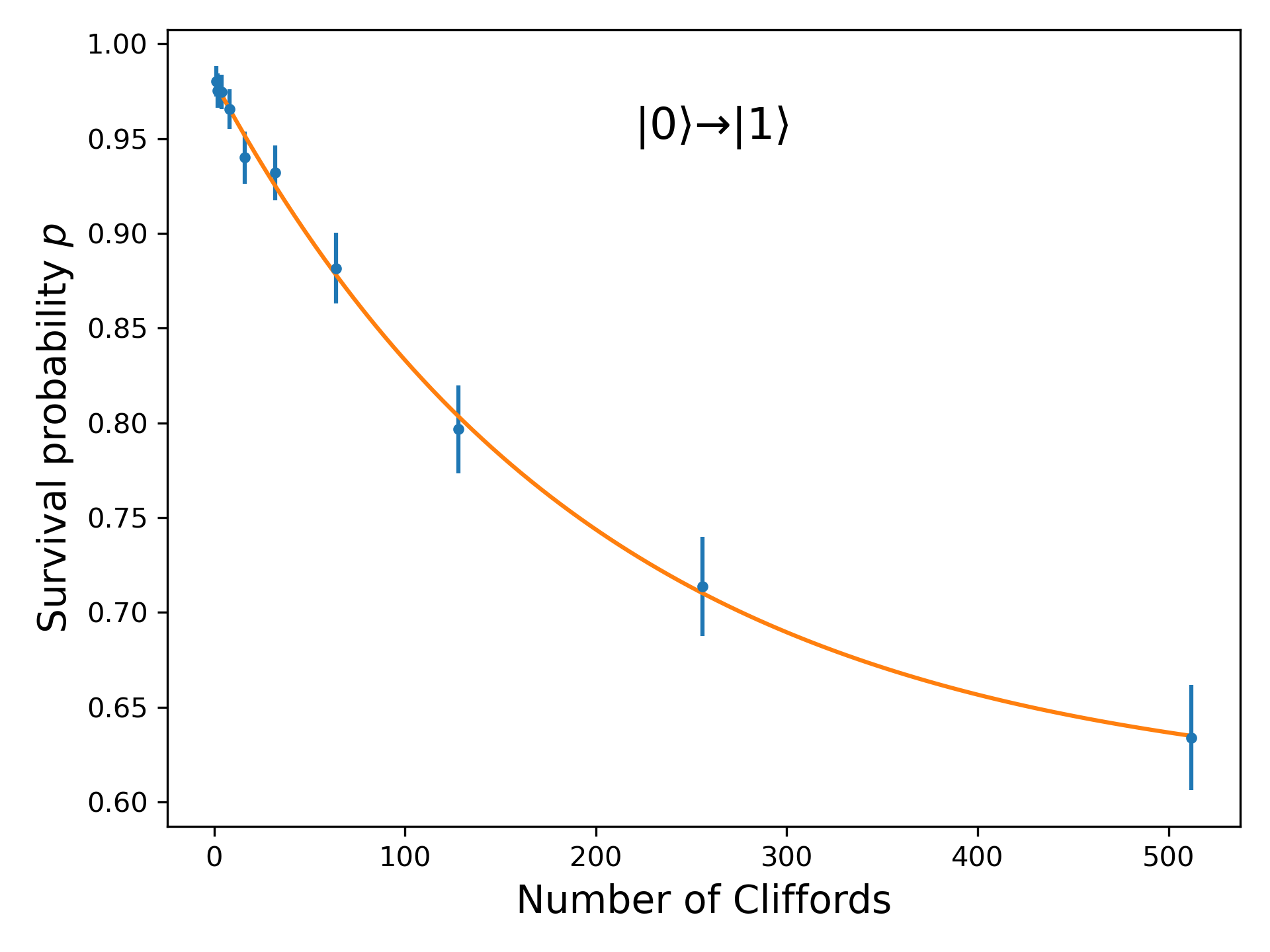}
}
\subfloat[\label{fig:SQ_RB_01}]{%
    \includegraphics[width=0.49\textwidth]{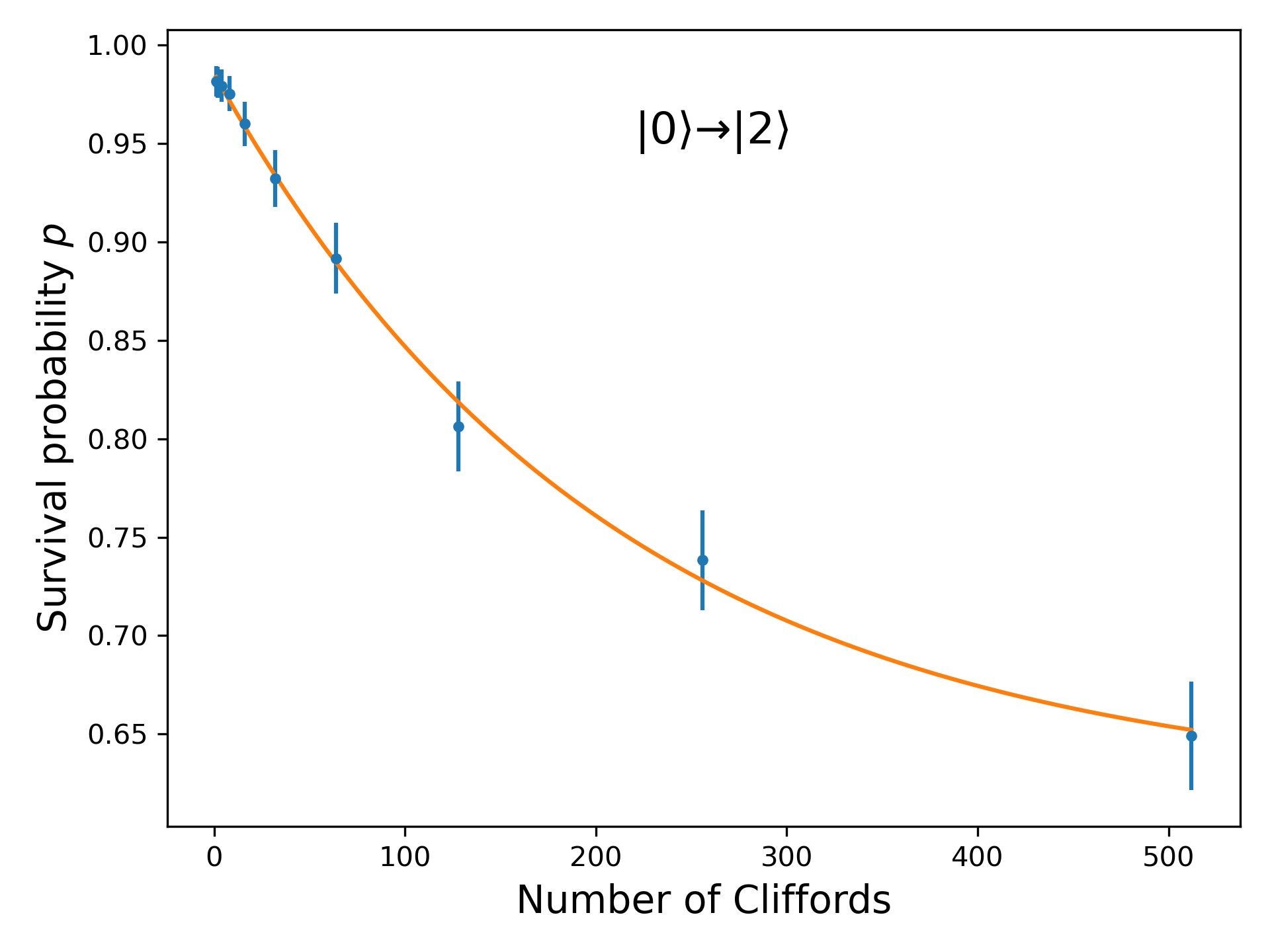}
}
\caption{Typical randomized benchmarking results on (a) $|0\rangle\to|1\rangle$ and (b) $|0\rangle\to|2\rangle$ transitions on a single ion in a chain of 8 particles. For each circuit length 10 random samples were generated, each measured 300 times. Error bars show statistical uncertainty. The solid line shows data fit with a $A + Bp^l$ function, where $l$ is the circuit length. Values of (a) $p = 0.9950 \pm 0.0011$ and $F_{SQ} = 0.9975 \pm 0.0006$ and (b) $p = 0.9952 \pm 0.0011$ and $F_{SQ} = 0.9976 \pm 0.0006$ are extracted from the fitting.}
\label{fig:SQ_RB}
\end{figure*}

We perform measurements on all 8 ions in the chain and all three qudit transitions. From ion to ion results differ only statistically. 
The difference between magnetically-sensitive and magnetically-insensitive transitions also appeared to be less than a statistical uncertainty. 
We present typical RB results in Fig.~\ref{fig:SQ_RB}. 
The mean single-qudit fidelity is extracted by fitting the $|0\rangle$ state survival probability dependence on the circuit length $l$ with a $A + Bp^l$ function, where $A$, $B$, and $p$ are fitting parameters. 
The fidelity given by $p+(1-p)/2$ is $99.75\pm 0.06\%$. {Our further investigations has shown that the leading contribution to the single-qudit gate infidelity is the addressing laser high-frequency phase noise.}

\subsection{Two-qudit gate benchmarking}

\begin{figure*}
\centering
\subfloat[\label{fig:2Q_Rabi}]{%
    \includegraphics[width=0.49\textwidth]{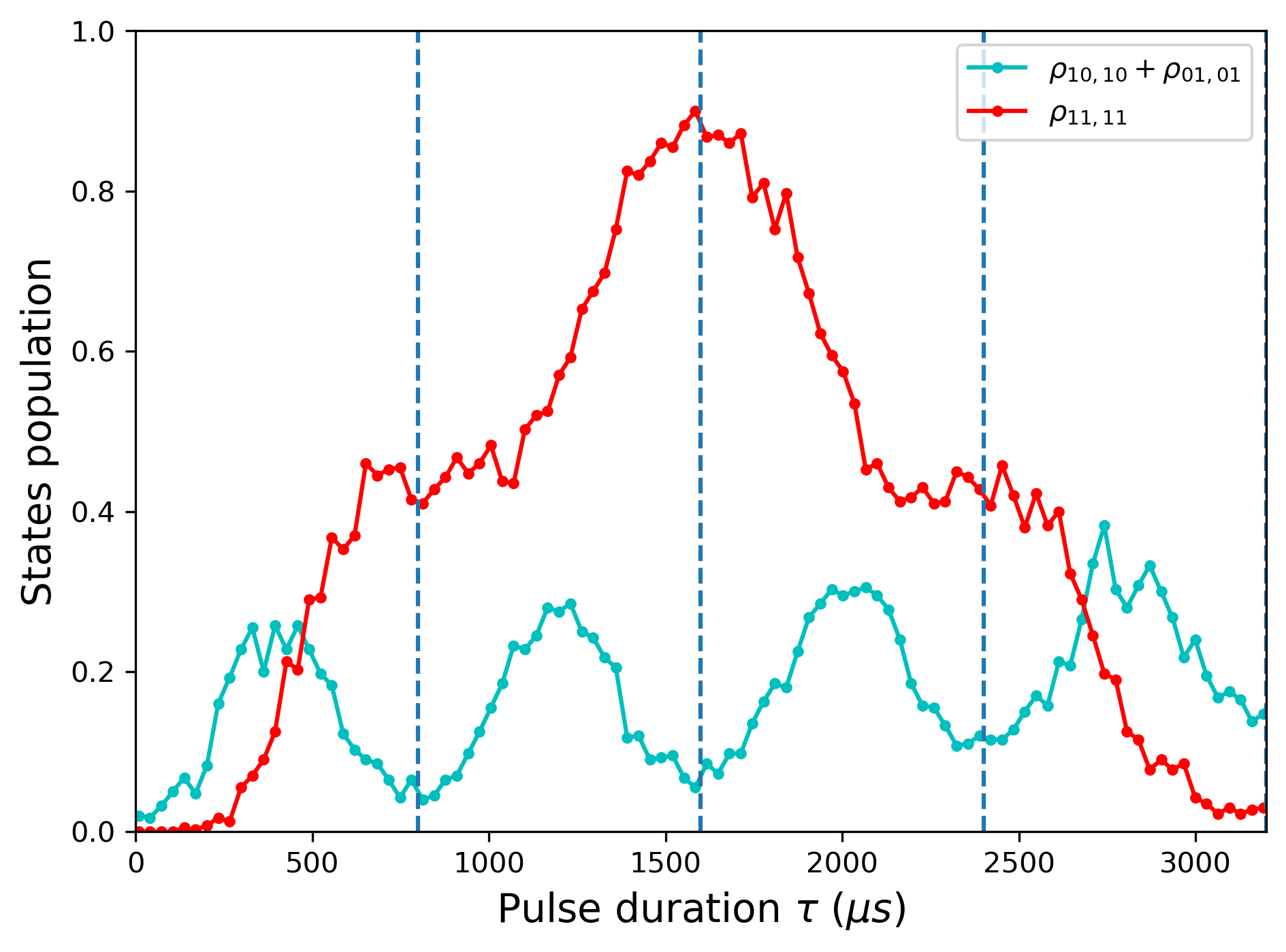}
}
\subfloat[\label{fig:2Q_parity}]{%
    \includegraphics[width=0.49\textwidth]{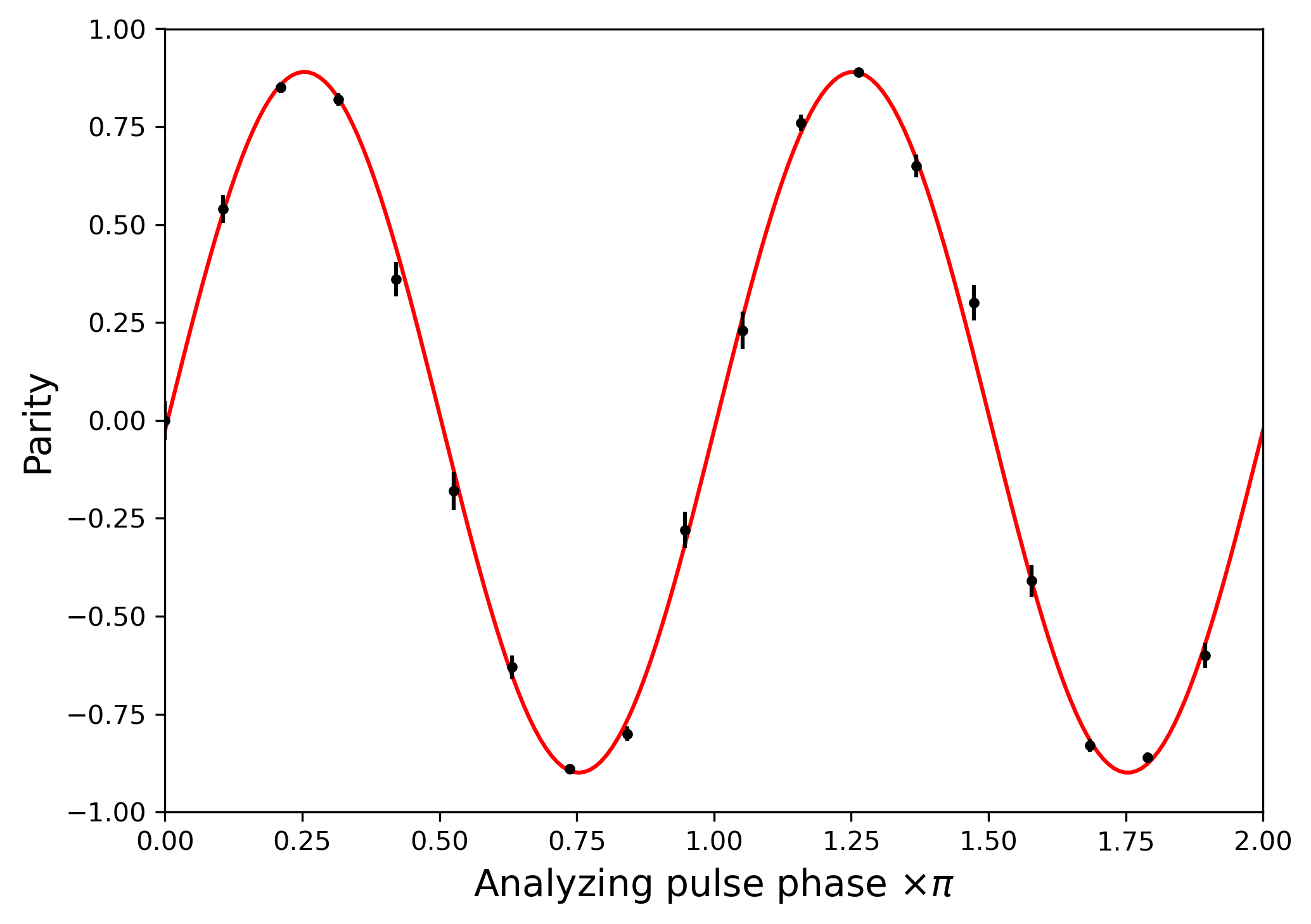}
}
\caption{Two-qudit gate benchmarking results between a pair of ions in the chain of 8 particles. (a) Population of the state $|11\rangle$ (red) and total population in states $|01\rangle$ and $|10\rangle$ (blue) as the function of the the entangling pulse length is shown. Vertical dotted lines show durations corresponding to one, two, three or four successive gates application. Total population is states $|00\rangle$ and $|11\rangle$ is $A = 0.96 \pm 0.01$. (b) Parity oscillations measured after successive application of the entangling gate and an analyzing $\pi/2$ pulse with varied phase $\phi$. The parity amplitude extracted from the weighted fitting of the data is $B = 0.895 \pm 0.005$.}
\label{fig:2Q}
\end{figure*}

To estimate fidelity of our two-qudit gate, we follow the method of Ref.~\cite{Blatt2008} and measure the preparation fidelity of the state $(|00\rangle-\imath|11\rangle)/\sqrt{2}$ from $|00\rangle$ by applying an $XX(\pi/4)$ gate to it. 
The fidelity can be estimated by measuring the total population $A$ in states $|00\rangle$ and $|11\rangle$ after application of the gate and parity oscillations contrast $B$. 
To measure the parity oscillations, we apply single-qudit gates $R^{01}_\phi(\pi/2)$ to both entangled ions and scan axis direction $\phi$. 
The parity is defined as $P=\rho_{11,11}+\rho_{00,00}-\rho_{10,10}-\rho_{01,01}$, where $\rho$ is a density matrix at the end of the circuit. 
The overall fidelity can be estimated as $F=A/2+B/2$. 
We show results of an entangling gate between a pair of ions when all 8 ions are trapped in Fig.~\ref{fig:2Q}. 
We reach two-qudit gate fidelities of $92.7 \pm 0.7\%$ including state preparation and measurement (SPAM) errors.

{The main contributions to this value are the SPAM error, a high-frequency phase noise of the addressing laser, long gate duration with respect to the coherence time (the two-qudit gate duration is 800 \textmu s, while the $T^*_2=16$ ms). Additional contributions come from system parameters drifts, such as drifts of secular frequencies, laser intensity, beam pointing errors, imperfect pulse shaping and imperfect cooling of the ions.  
}

\subsection{Coherence time}

$T_1\approx53$~ms in our system is determined by the spontaneous decay from the states $|k\rangle$, $k>0$. 
If the magnetic field noise is suppressed only with an active feedback loop and experiment line-triggering, we achieve $T^*_2=16$~ms on a magnetically-insensitive $|0\rangle\to|1\rangle$ transition and only $T^*_2=1$~ms on other qudit transitions (more details are given in Ref.~\cite{Zalivako2023}). The $T^*_2$ for the $|0\rangle\to|1\rangle$ is mainly determined by the laser frequency noise, as it does not depend on the bias magnetic field value and turning the line-triggering on and off. 
In case of magnetically-sensitive states a relatively short $T^*_2$ time is caused by magnetic field noise. To overcome this problem, we develop the schemes of continuous dynamical decoupling, which raised the magnetic-sensitive levels coherence time up to 9~ms~\cite{Zalivako2023}. 

\section{Realization of quantum algorithms}\label{sec:algorithms}

Another approach to a quantum processor benchmarking is to run quantum algorithms.
For example, Bernstein-Vazirani~\cite{Bernstein1997} and Grover's~\cite{Grover1996} algorithms are widely used for this purpose~\cite{Blatt2003,Blatt2016,Monroe2019}. Since the expected output state probability distributions after running these algorithms can be rather easily simulated classically, one may compare them with the output of the processor, as suggested in Ref.~\cite{Monroe2019}. 
This benchmarking technique is more comprehensive than the component-based approach, as it not only catches errors from all the system components but also takes into account their interactions. 

\subsection{Algorithmic benchmarking in the qubit regime}

{The first stage of algorithmic benchmarking was performed with the quantum processor operating in the qubit regime, namely when only two levels $\ket{0}$ and $\ket{1}$ were employed for the information encoding. In this regime, the processor uses the most of its native quantum gates and thus its performance can be benchmarked in many aspects, while the results can be directly compared to the other ion systems.}

{We note, that in this regime single- and two-qubit gates in the algorithms are directly transpiled to the corresponding single-qudit operations between levels $\ket{0}$ and $\ket{1}$ and native MS two-qudit operations.}

\subsubsection{Two-qubit Bernstein-Vazirani algorithm}

\begin{figure}
\centering
\includegraphics[width=0.48\textwidth]{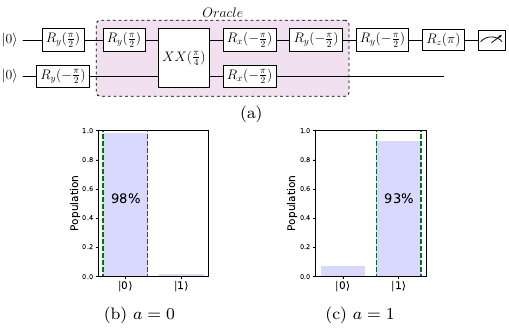}
\caption{Two-qubit Bernstein-Vazirani circuit with native ion gates. In (a) the oracle is given for $a = 1$. In case $a=0$ oracle is equivalent to the identity operator. Success probabilities for $a=0$ and $a=1$ are depicted on (b) and (c), correspondingly.}
\label{fig:bva1}
\end{figure}

Bernstein-Vazirani algorithm~\cite{Bernstein1997} deals with the class of Boolean functions $f(x):\{0,1\}^n \rightarrow \{0,1\}$, which is known to be the dot product between its argument $x=(x_0,\ldots,x_{n-1})$ and a secret bit string $a=(a_0,\ldots,a_{n-1})$:
\begin{equation}
    f(x) = x\cdot a = \sum_i x_i a_i~(\text{mod}~2).
\end{equation}
The task of the algorithm is to find this bit sting $a$. 
The quantum oracle for this problem has the form
\begin{equation}
    U_f:\ket{x}\otimes\ket{t}\to\ket{x}\otimes\ket{f(x)\oplus t}
\end{equation}
(here $\oplus$ denotes mod 2 summation).
The essence of the Bernstein-Vazirani algorithm is in obtaining the value of $a$ using a single query to the quantum oracle:
\begin{equation}
    ({\sf H}^{\otimes(n+1)} U_f {\sf H}^{\otimes (n+1)}) \ket{0}^{\otimes n}\otimes\ket{1} = \ket{a}\otimes\ket{1},
\end{equation}
where ${\sf H}$ denotes a standard Hadamard gate.

To execute the Bernstein-Vazirani algorithm on our trapped-ion processor, we transpile the circuit to the set of native single-qudit and two-qudit gates {between levels $\ket{0}$ and $\ket{1}$}. 
Each part of the algorithm (preparation of a uniform superposition of input states, oracle and answer decoding) is transpiled separately since all other parts of the algorithm should be independent from the oracle.
For this reason, we do not unite rotations from the oracle with nearby standing gates. 
As Bernstein-Vazirani algorithm requires one ancilla qubit, the key length of the two-qubit algorithm is one bit.
In the case of ideal implementation of the algorithm, the oracle secret key is given by the output state of the first qubit with 100\% success probability.
The transpiled circuit and experimental results  are presented in Fig.~\ref{fig:bva1}.
Both circuits were executed 1024 times.
The measured success probability averaged over secret key values is 95\%.

\subsubsection{Two-qubit Grover's algorithm}

Grover's algorithm~\cite{Grover1996,Grover1997} considers a black box function $f(x): \{0,1\}\rightarrow\{0,1\}$, which yields 0 for all bit strings except one ($s$), for which the output is 1 ($f(s)=1$). 
Grover’s algorithm finds this special value $s$. The function $f(x)$ is encoded in the oracle.
The algorithm creates a uniform superposition of all possible input states, which is followed by several cycles of Grover iterations, each resulting in the amplification of the $|s\rangle$ amplitude. 
Grover algorithm provides a quadratic speedup, however, it has been proven~\cite{Bennett1997} that no higher speedup is possible for this problem.

\begin{figure*}
\centering
\includegraphics[width=0.9\textwidth]{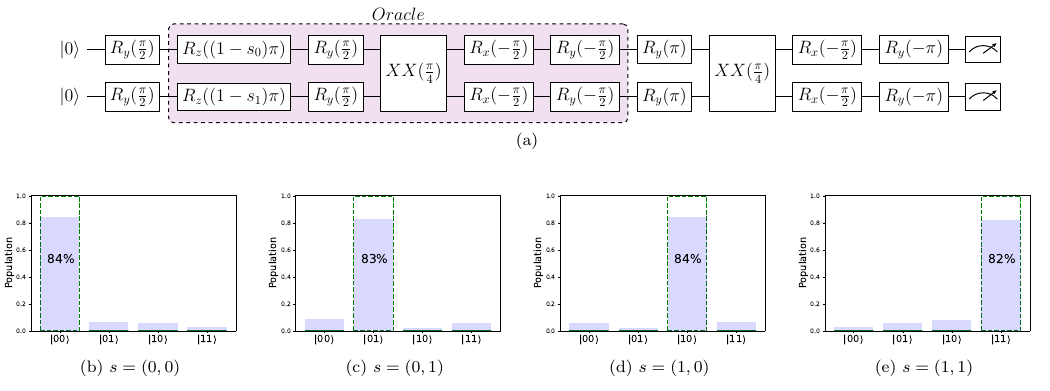}
\caption{In (a) transpiled Grover's algorithm circuit for $s=(s0,s1)$ is presented. Experimental results of its execution for all possible values of $s$ are depicted in (b-e).}
\label{fig:grover}
\end{figure*}

The single iteration of the Grover algorithm consists of two parts: the oracle $U_f\ket{x}\mapsto (-1)^{f(x)}\ket{x}$, which applies the phase factor $(-1)^{f(x)}$ to the state $\ket{x}$, and a diffusion operator, which inverts the state around the average.
Here we consider a two-qubit version of Grover's algorithm for 4 possible variants of the value $s$ and keys of the oracle, respectively.
For the two-qubit case, it is sufficient to implement only one iteration of the algorithm, which includes two two-qudit gates: one in the oracle and one in the diffusion operator. 
The transpiled circuit for this algorithm and the experimental results of its execution for different values of $s$ are presented in Fig. \ref{fig:grover}.
Each circuit has been executed 1024 times.
The average success probability of algorithm execution is 83\%.

\subsubsection{Quantum chemistry: H$_2$ and LiH simulations} 

\begin{figure*}
\centering
\subfloat[\label{fig:subfirst}]{%
    \includegraphics[width=0.49\textwidth]{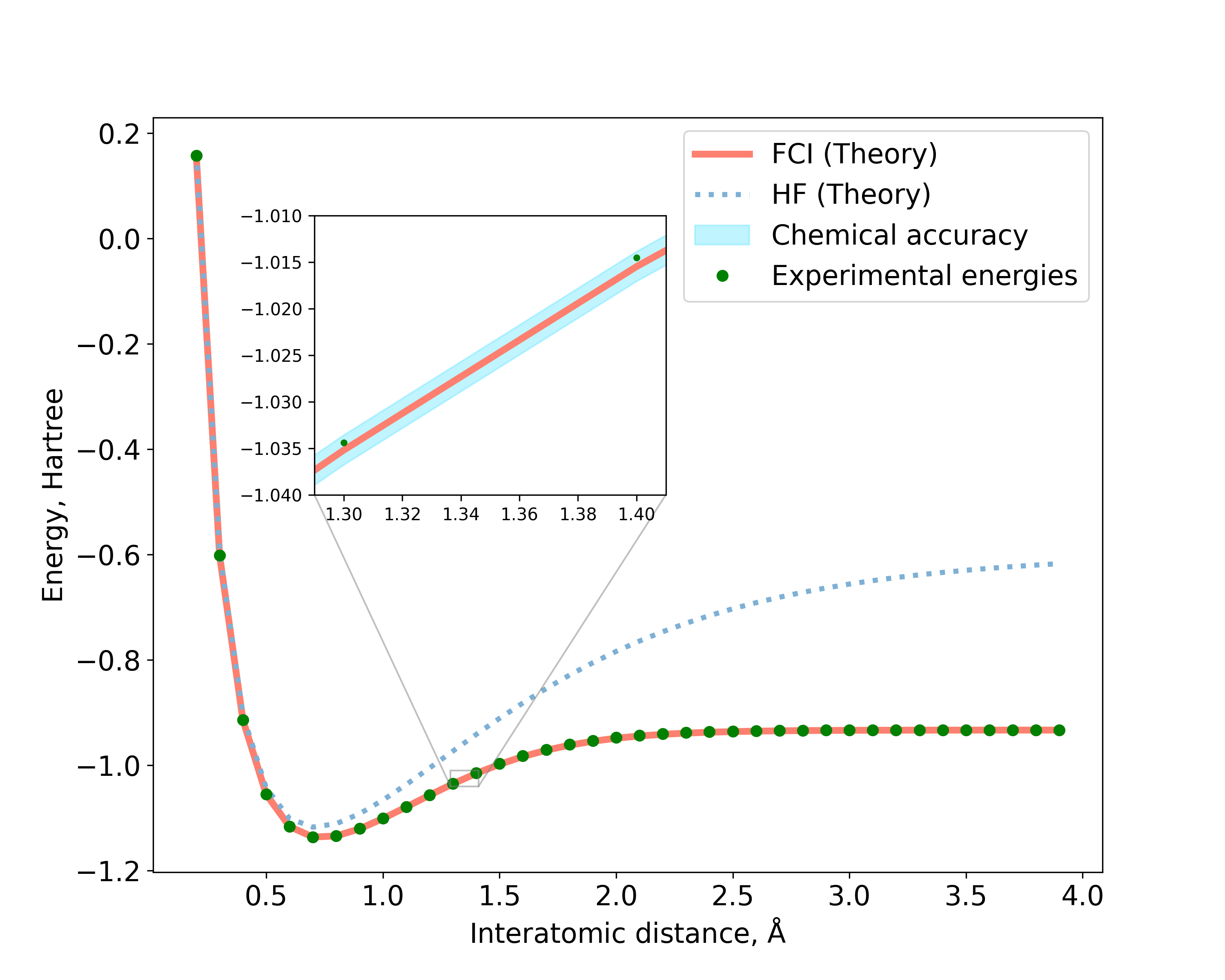}
}
\subfloat[\label{fig:subsecond}]{%
    \includegraphics[width=0.49\textwidth]{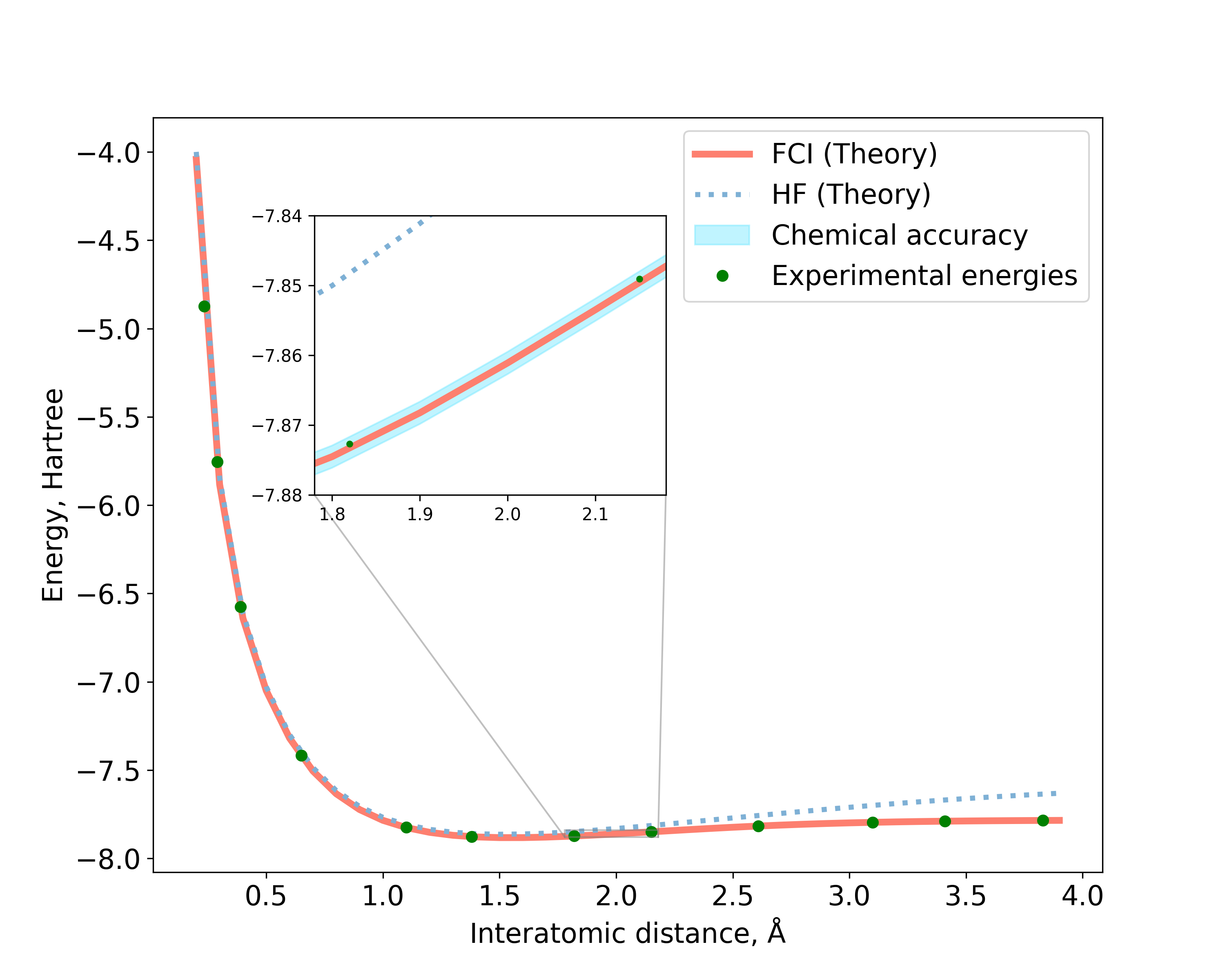}
}
\caption{(a) Potential energy surface for $\mathrm{H}_2$ molecule. The blue-shaded region around the Full Configuration Interaction (FCI) curve in the inset represents the chemical accuracy threshold of $0.0016$ Hartree. Starting with state $|01\rangle$, we used the Hardware-Efficient circuit $C(\boldsymbol{\theta}) = R_{y}^{0}(\theta_0)R_{y}^{1}(\theta_0)CX(0,1)R_{y}^{0}(\theta_1)R_{y}^{1}(\theta_1)$ with fixed parameters $\theta_0 = 0.5477$ and $\theta_1 = 0.0703$ and 4096 shots. The upper index of the rotation gate corresponds to the qubit index. (b) Potential energy surface for $\mathrm{LiH}$ molecule. The blue-shaded region around the FCI curve in the inset represents the chemical accuracy threshold of $0.0016$ Hartree. We use $|10010\rangle$ as the initial state, which is supplemented with a parameterized Hardware-Efficient circuit having 10 fixed parameters, taking 512 shots for each circuit execution.}
\label{fig:figures}
\end{figure*}

We follow the Iterative Quantum Assisted Eigensolver(IQAE)~\cite{iqae, qas, general_qas, sdp} algorithm for computing molecular ground state energies (see Suppl. materials).
Under the Born-Oppenheimer approximation, the molecular Hamiltonian is usually expressed in its second-quantized form: 
\begin{equation}\label{chem_ham}
    H = \sum_{p,q=1}^{N}h_{pq}a_{p}^{\dagger}a_{q} + \frac{1}{2}\sum_{p,q,r,s=1}^{N}g_{pqrs}a^{\dagger}_{p}a^{\dagger}_{r}a_{s}a_{q},
\end{equation}
where $a^{\dagger}_{p}(a_{p})$ is the fermionic creation (annihilation) operator and $N$ is the number of molecular basis functions. 
The coefficients $h_{pq}$ and $g_{pqrs}$ are called one- and two-electron integrals and can be computed classically. 

This Hamiltonian is transformed into a {weighted} sum of qubit operators, or Pauli strings, which are tensor products of Pauli matrices $X$, $Y$, $Z$, and the identity operator $I$. After the transformation, the Hamiltonian is expressed as $H = \sum_{i} \beta_i U_i$, where each $U_i$ represents a Pauli string.

For our molecular simulations of both H$_2$ and LiH, we use the minimal basis set STO-3G~\cite{Hehre_sto3g} (Slater Type Orbital {approximated by three} Gaussian functions). 
Specifically, for the H$_2$ molecule, we employ the Parity~\cite{parity} transformation with a two-qubit reduction to derive a two-qubit Hamiltonian. 
The use of an order of Krylov subspace $K=1$ (see Suppl. materials) is sufficient to achieve results within the chemical accuracy range, 
resulting in overlap matrices of dimension $5 \times 5$.
The results for H$_2$ are shown in Fig.~\ref{fig:subfirst}.

For the LiH molecule, we use Jordan-Wigner~\cite{jw} transformation along with the qubit-tapering procedure~\cite{tapering} and froze orbitals to obtain a 5-qubit Hamiltonian for our simulation. 
Again, by using an order of Krylov subspace $K=4$, we can achieve results within the chemical accuracy range, resulting in overlap matrices of dimension $1024 \times 1024$. 
The results for the LiH molecule are presented in Fig.~\ref{fig:subsecond}.

\subsection{Algorithmic benchmarking in the ququart regime}

{While the qubit regime allows one to benchmark the majority of the processor features and subsystems, such as single- and two-qudit gates, which are common for both regimes, individual addressing and the influence of the cross-talks, readout errors, decoherence due to ions spontaneous decay, laser instabilities and ion string heating, some of the noise sources in the qudit processor are unique to the qudit regime. The most significant of such error sources are the relative coherence between different qudit levels and effect of the quantum gates on the spectator levels. The first of these sources can be relevant as various qudit levels have different sensitivity to the magnetic field fluctuations, while the second one can occur due to Stark shifts on the spectator levels. The simplest algorithm which can capture both of these effects is a two-qubit Bernstein-Vazirani algorithm transpiled to be run within a ququart. In this transpilation it has a form similar to the Ramsey-type experiment. It is inherintly sensitive to the coherence between levels as well as to the both phase and population changes in spectator states during single-qudit operations.}

We encode qubits in a qudit in the following way:  
   $\ket{0} = \ket{01}$,
    $\ket{1} = \ket{11}$,
    $\ket{2} = \ket{10}$,
    $\ket{3} = \ket{00}$.

\begin{figure}
\includegraphics[width=0.48\textwidth]{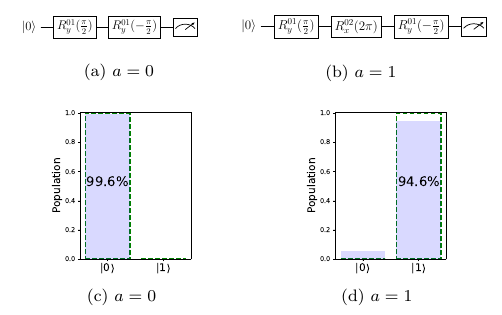}
\caption{Single-qudit Bernstein-Vazirani circuits for chosen qubit-to-ququart encoding for $a = 0$ and $a=1$ (on (a) and (b) respectively) and experimental results of their execution (on (c) and (d)).
}
\label{fig:bv_qudit}
\def\rx#1{\gate[][1cm]{R_x(#1)}}
\def\ry#1{\gate[][1cm]{R_y(#1)}}
\def\rz#1{\gate[][1cm]{R_z(#1)}}
\end{figure}

The transpiled circuit and experimental results are presented in Fig. \ref{fig:bv_qudit}. 
Each circuit was executed 1024 times {ending with a full ququart readout procedure.} 
The measured success probability averaged over secret key values is 97\%, {which proves coherence between levels and} exceeds the result for the {qubit regime. This is an expected result as the two-particle entangling operation in the algorithm is replaced with a single-qudit one, which is one of the features of the qudits, resulting in a better process fidelity.}
The main contribution to the error here is a {SPAM error. This experiment showed no evidences on the cross-talk between single-qudit operations acting on different level pairs or additional phases acquired by the spectator states due to Stark effect.}

\section{Conclusion}\label{sec:conclusions}

In this work, we have presented the 8-ion-based qudit quantum processor. 
The developed qudit processor uses trapped $^{171}$Yb$^{+}$ ions, where quantum information is encoded in Zeeman sublevels of states $^2S_{1/2}(F=0)$ and $^2D_{3/2}(F=2)$, coupled by an E2 transition at 435~nm. Using four states in each particle for information encoding (ququarts) makes the presented setup computationally equivalent to a 16-qubit quantum processor. We described details of the setup, including a new Paul trap, addressing and readout systems, which provided significant improvements over the previous generation of the experiment~\cite{Kolachevsky2022} and allowed running quantum algorithms. We presented the results of component-based system benchmarking, including fidelities of single- and two-qudit operations. At the same time, we started algorithmic benchmarking of our processor, gradually increasing the complexity of the problems. We showed results of exectuing two-qubit Bernstein-Vazirani and Grover's search algorithms, as well as H$_2$ and LiH molecular simulations. {We have also compared the performance of the two-qubit Bernstein-Vazirani algorithm in qubit and ququart regimes, showing that embedding several qubits inside one particle can give an advantage in the algorithms performance.  This is an important step towards efficient exploiting properties of multilevel systems for useful applications. Despite more complicated readout procedure and higher sensitivity to the various decoherence sources the qudit approach enables one to at least double the computational space dimension with the same number of particles as well as run some algorithms more efficiently due to replacing multiparticle gates with single-qudit ones and reducing number of entanglements required~\cite{Nikolaeva2023}.}
Our future plans include both further studying of our system by running more complicated algorithms, utilizing larger Hilbert space, available in our system, and continuing improving the system performance. The former is also in line with experimental studying of advantages of the qudit approach in comparison with conventional qubits, in particular, for Grover's search algorithm~\cite{Nikolaeva2023}. 
To improve the system performance, we are actively working on the reduction of the addressing laser noise, magnetic field fluctuations and more efficient experiment control protocols.

\section*{Acknowledgements}

This work was supported by Rosatom in the framework of the Roadmap for Quantum computing (Contract No. 868-1.3-15/15-2021 dated October 5).

\section*{Conflict of Interest}
The authors declare no conflict of interest.

\section*{Data availability}
The data that support the findings of this study are available from the corresponding author upon reasonable request.

\bibliography{bibliography.bib}

\end{document}


\section*{Supplemental materials}
\subsection{Methods}
\subsubsection{Linear Paul trap}

The ion trap is placed in the stainless steel vacuum chamber, where the vacuum is maintained using an ion-getter pump. 
After high-temperature baking, the pressure in the chamber is below $10^{-10}$~mbar. 
We expect the actual pressure to be lower according to the observed ion's lifetime in the trap.  

The trap consists of four blade-like electrodes, which provide a radial confinement and two cylindrical end-cap electrodes for the axial confinement. 
The holes are drilled in the center of the end-cap electrodes to provide axial optical access. 
The trap electrodes are made of gold-plated oxygen-free copper, while the electrical insulation between them is made of alumina dielectrics. 
The distance between the blades and the trap center is $r_0=250\mu m$. 
The trap is assembled on the custom-made flange with all required electrical feed-throughs. 
The trap can be cooled by refrigerant supplied by an additional copper tube feedthrough. 
It enables cooling the trap down to $-100\,^\circ$C by pumping a temperature-stabilized cold gas or liquid through the tube. 
Cooling the trap reduces the ion heating rate and improves vacuum conditions. 
Besides cooling, this provides the trap temperature stabilization during its operation, which is important for the stability of the radial secular frequencies. 
An atomic oven with a natural abundance of ytterbium isotopes is attached to the trap mount to assist in loading ions. 

\begin{figure}[!htb]
\center{\includegraphics[width=0.5\linewidth]{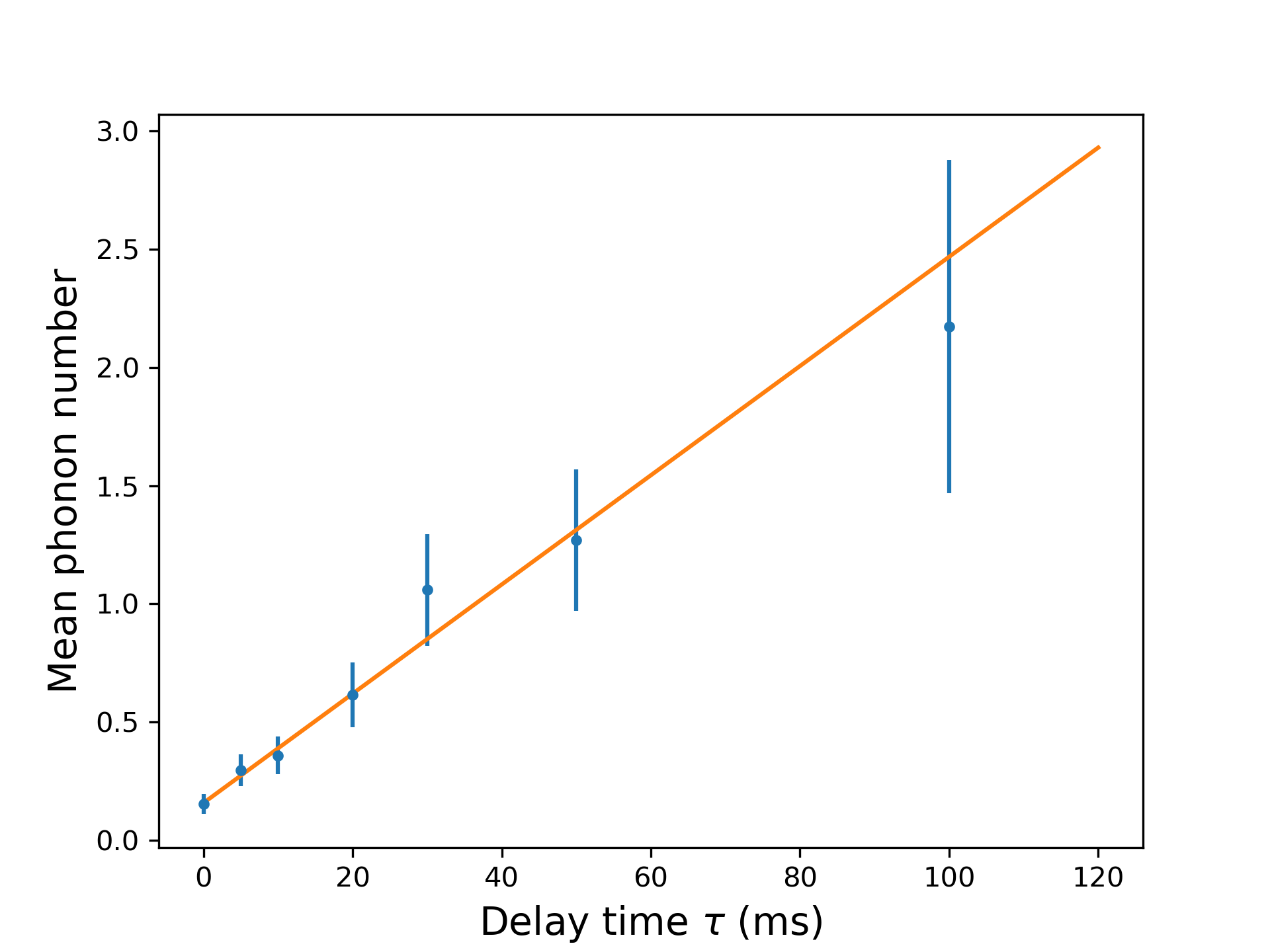}}
	\vskip-3mm
\caption{Results of the heating rate measurement along the $x$-axis (radial) of the trap. The single ion inside the trap was ground-state cooled and after a varied delay time $\tau$ a mean phonon number in the mode has been measured by the ratio of the red and blue secular sidebands~\cite{Leibfried2003}. Each point is an average of 300 experimental shots. Solid line shows a linear approximation of the data yielding a heating rate of $\dot{n}=23 \pm 3$~phonons/s. The trap was operated at secular frequencies $\{\omega_x, \omega_y, \omega_z\} = 2\pi \times \{3.7, 3.8, 0.116\}$~MHz. }
\label{fig:heating_rate}
\end{figure}

Radiofrequency (RF) trapping potentials are applied to the trap electrodes via a bifilar helical resonance transformer. 
It provides the possibility to apply independent DC voltage offsets to all four blades. These four potentials provide enough degrees of freedom to both compensate for an excess micromotion and to introduce asymmetry between radial trap axes ($x$ and $y$). 
The latter is important for efficient Doppler cooling and spectral separation of $x$ and $y$ modes during entangling operations. 
The maximum achieved radial secular frequency with this setup for a single $^{171}$Yb$^{+}$ was $2\pi\times4.4$~MHz at the trap drive frequency of $\omega_{RF}=2\pi\times 30.8$~MHz and the input RF power below 5~W. 
The operational parameters for the majority of the experiments described below were $\{\omega_x, \omega_y, \omega_z\} = 2\pi \times \{3.7, 3.8, 0.116\}$~MHz for a single $^{171}$Yb$^{+}$ ion. 
The measured heating rate in the radial direction at these parameters is $\dot{n}=23 \pm 3$~phonons/s, see Fig.~\ref{fig:heating_rate}. 

The radial secular frequencies are actively stabilized in a similar way to the Ref.~\cite{johnson2016active}. 
The amplitude of the radiofrequency potential, which determines the secular frequencies, is sampled with a capacitor divider right before the vacuum chamber, rectified, and fed to a servo loop controlling the amplitude of the generator signal seeding the resonance transformer. 
Given the temperature stability in the laboratory within the range of 0.2~K, the radial secular frequency fluctuations below 200~Hz were measured during the 12~hours interval.

Several viewports provide optical access to the ions. 
One of the access directions is along the trap axis $z$, see Fig.~2 in the main text. 
It is used for repumping beams at 935~nm and 760~nm (more details on the laser systems and experimental protocols are given in the next sections), as well as for an auxiliary Doppler cooling and readout beam at 369~nm. The main Doppler cooling beam at 369~nm and a photoionization beam at 399~nm are focused to the trap center at $45^\circ$ angle to its axis. 
All these beams do not require tight focusing, so these viewports have small numerical apertures (NA). A large window with $NA=0.48$ is used to collect the ions' fluorescence to read out qudits states. The collimated laser beams at 435~nm which are used to perform single- and two-qudit gates enter the the vacuum chamber orthogonal to the trap axis and are focused to the ions using an in-vacuum composite lens with $NA=0.2$ (individual addressing beams, IAB). 
Another weakly focused addressing beam at 435~nm illuminating all ions simultaneously is orthogonal both to the trap axis $z$ and the IABs. It is used for micromotion detection and compensation. 
All laser radiation, except IABs, is delivered to the vacuum chamber by optical fibers. 

To avoid coherent population trapping during Doppler cooling and readout~\cite{Berkeland2002, Ejtemaee2010} as well as to lift the degeneracy between $^2D_{3/2}$ qudit states, the magnetic field of $B=6.1$~G is applied inside the trap using permanent neodymium magnets. Its direction is orthogonal to the trap axis $z$ and has an angle of approximately $60^\circ$ to the individual addressing beams direction, which allows driving transitions between $|0\rangle$ and all other qudit states. An additional set of three pairs of coils in the Helmholtz configuration is used to actively stabilize the magnetic field. 
A three-axis magnetic field sensor is installed approximately at 4 cm from the trap center.

\subsubsection{Trap loading}

To load ions we use an atomic oven and two photoionization laser beams focused on the trap center. The first laser (external cavity diode laser, ECDL) is resonant to the $^1S_0 \to \,^1P_1$ transition at 399~nm and isotope-selectively excites neutral ytterbium atoms to an intermediate $P$ state~\cite{Kleinert2016}. 
The second laser is a Doppler-cooling beam at 369 nm, which ionizes atoms from this state to a continuum. 
Parameters of the 399~nm beam and atomic oven are set in such a way, that a single $^{171}$Yb$^{+}$ ion is loaded every minute at average. 
The number of loaded ions is monitored during the process by their fluorescence signal with a high-sensitive camera. 
Using this standard approach we can deterministically load ion strings of various lengths in the trap.

\subsubsection{Doppler cooling and state initialization}

The Doppler cooling of the ions is performed on a quasi-cyclic $^2S_{1/2}(F=1)\to \,^2P_{1/2}(F=0)$ transition at 369~nm with the natural linewidth of $\Gamma=2\pi\times20$~MHz. 
The source is the fiber laser, whose frequency is tripled via nonlinear processes to reach the desired wavelength. 
The fiber laser not only ensures better frequency stability in comparison to the ECDLs at this wavelength, but also provides higher optical power. 
During the cooling process, the laser beam is phase-modulated at 14.7~GHz with a free-space electro-optic modulator (EOM) to repump population from another hyperfine component of the ground state $^2S_{1/2}(F=0)$ by driving $^2S_{1/2}(F=0)\to \,^2P_{1/2}(F=1)$ transition. 
Another free-space EOM in the beam path enables modulation of the beam at 2.1~GHz, which is necessary for the state initialization (this procedure is described below). 
The beam is then split into two parts, which amplitudes and frequencies can be individually controlled with acousto-optical modulators (AOMs). Both beams are sent via single-mode optical fibers to the vacuum chamber. The first beam is the main cooling beam and has a non-zero projection on all three principal axes of the trap.
The second beam propagates along the trap axis $z$ and assists faster recrystallization of the ions in the case of the crystal melting (due to collisions with the background gas). 
It is also used for the readout process causing state-dependent ion fluorescence. The second beam is turned on only in between of experimental shots and during readout. 
As both Doppler cooling and readout efficiencies strongly depend on the cooling beam intensities, their optical powers are actively stabilized.

As the $^2P_{1/2}(F=0)$ state has 0.5~\% probability to decay in the metastable $^2D_{3/2}(F=1)$ state, an additional repumping beam at 935~nm is also used during the cooling process to quench the $^2D_{3/2}$ state population back to the ground state via an auxiliary $^3[3/2]_{1/2}$ level. 
The corresponding transition has the natural linewidth of $4.2$~MHz. 
The beam source at 935~nm is a distributed feedback (DFB) diode laser. 
It is fiber-coupled and passed through a fiber EOM, providing its phase-modulation at 3.07~GHz. 
The EOM helps to clear the population from the $^2D_{3/2}(F=2)$ hyperfine sublevel as well (Fig.~1 in the main text), which we use to reset the qudit state. 
The frequency and the amplitude of the beam are controlled with a fiber AOM. 
It also allows us to actively stabilize the beam intensity right before entering the vacuum chamber.

Both laser sources are frequency-stabilized using a multichannel wavemeter with a built-in proportional-integral-differential (PID) controller. 
The wavemeter is regularly calibrated with respect to the addressing laser fundamental wavelength at 871~nm, thus ensuring the frequency stability better than $\pm2$~MHz near its calibration point.

We also have another repumper laser in our system at 760~nm which helps to quench the metastable $^2F_{7/2}$ state sometimes populated due to collisions with the background gas~\cite{Hoang2020}. This is an another DFB diode laser, which is phase modulated at 5.2~GHz with a free-space EOM to clear all $^2F_{7/2}$ hyperfine components. Its frequency is also modulated at a 10~Hz rate with amplitude of 100~MHz to depopulate all Zeeman components. The laser light is delivered to the trap via an optical fiber and can be switched on and off with a mechanical shutter.

The Doppler cooling process takes place at the beginning of each experimental shot and takes 6~ms. During that period both the main cooling beam (369~nm) and the repumping beam (935~nm) are on as well as the EOMs at 14.7~GHz (in 369~nm beam) and 3.07~GHz (in 935~nm beam). The cooling light is red-detuned from the transition resonance by $\Gamma/2$ and its intensity is set to be equal to the saturation intensity. The repumping beam is set to the transition resonance and is operated well above saturation. The beams' polarizations are aligned to maximize the ion's fluorescence~\cite{Ejtemaee2019}. 
The ion's temperature after the cooling cycle, which is measured using the Rabi oscillations damping method~\cite{semenin2022determination}, is 1.6(1)~mK.

After the Doppler cooling, the ions are initialized to the $|0\rangle=\,^2S_{1/2}(F=0,m_F=0)$ state by switching off the 14.7~GHz EOM and replacing it with a 2.1~GHz EOM in the main cooling beam at 369~nm. Such field configuration causes optical pumping to the $|0\rangle$. The intensity of the cooling beam during this process is also raised above the saturation to speed up the process, which takes $2\mu s$. 

\subsubsection{Ions addressing laser system}

The addressing laser is one of the most crucial components in the system, which is aimed to perform high-fidelity control of the optical ion qudits. 
The laser source must feature both high mid- and long-term frequency stability to ensure long coherence times. 
At the same time, it should possess  the lowest possible  phase noise at detunings  from 100~kHz to 5~MHz (relative to the carrier), because the  fidelity of  single- and two-qudit gates is extremely sensitive to its presence~\cite{nakav2023effect, Kolachevsky2022}. 
The laser beam quality, high spatial resolution and mechanical stability are also necessary for individual ion addressing in the chain with low cross-talks. 
Prompt and accurate switching between different ions is also crucial for achieving the full control over the system.

Our laser system is based on an ECDL at 871~nm, which is frequency locked using Pound-Drever-Hall (PDH) technique~\cite{Salomon1988} to a home-built ultrastable cavity made from ultra-low expansion (ULE) glass. The details on the cavity and the locking scheme can be found elsewhere~\cite{Zalivako2020}. The system demonstrates relative frequency instability below $3\times10^{-15}$ in the range of averaging times between 0.5~s and 50~s. Unfortunately, diode lasers are known for excessive level of high-frequency noise, which results in prominent peaks in the phase noise density spectra at the detunings close to the servo loop bandwidth (usually around 1~MHz, so-called ``servo-bumps''). This high-frequency phase noise significantly impacts fidelity of two-qudit operations in our setup~\cite{Zalivako2020}. At the moment, we investigate several approaches to filter out this noise; however, in the experiments described below no filtering is used.

The 871\,nm radiation is amplified in a tapered amplifier and is frequency doubled in a non-linear crystal inside a bow-tie cavity. The available laser power at 435~nm is 1.3~W. 
The laser light is divided into three parts, each passing its own AOM for frequency, phase, and amplitude control. 

Two of these beams are used for individual addressing (IABs). The direction of both beams is controlled with acousto-optical deflectors (AODs), which enables us to scan the beams along the ion chain. These beams are combined on a beamsplitter, expanded, and focused on the trap center using in-vacuum lens. Special care was taken to provide mechanical stability of all optical elements and prevent beam-pointing errors. The whole beam path has been enclosed into an insulating box to reduce the effects of the air flows.

Fine alignment of IABs is performed with piezo-driven mirrors in the beam path. During the alignment procedure, the piezo voltages for each beam are scanned and ion excitation probabilities are registered, visualizing the actual beam profile. The beams' waist at the ion's position  was measured to be 2~$\mu m$ (full width on the half maximum) and is mostly determined by misalignment of the in-vacuum lens. 

At {usual} trapping conditions ($\omega_z=2\pi\times116$~kHz and 8 ions), the minimal distance between neighboring particles is 7.3~$\mu m$. The addressing cross-talks were measured by addressing a single ion in the chain and measuring the Rabi flopping on the $|0\rangle\to|1\rangle$ transition. The cross-talks, defined as a ratio of the Rabi frequency on the spectator ion to the one on the target particle, were estimated to be 3-5\% depending on the ion.

The last of three beams at 435~nm is used for ion's global addressing orthogonally to the IABs. 
It is passed through another AOM to compensate for a frequency shift in AODs in IABs and is sent to the trap by an optical fiber. 

\subsubsection{Micromotion compensation}

We detect excess micromotion~\cite{Berkeland1998,Leibfried2003} in the radial plane by measuring the ratio of Rabi frequencies on the carrier transition $|0\rangle\to|1\rangle$ and its micromotion sideband. 
As we have two orthogonal addressing beams in this plane (IABs and the global addressing beam), it is possible to extract the information about micromotion along both of these axes. 
With the use of this detection technique, we minimize radial micromotion by adjusting DC potentials on the trap blade electrodes.

\subsubsection{Ground state cooling}

Although the M\o{}lmer-S\o{}rensen two-qudit gate does not necessarily requires ground state cooling~\cite{Molmer-Sorensen1999-2}, the thermal motion at Doppler limit temperature still decreases its fidelity~\cite{haljan2005spin}. 
Therefore, we use the resolved-sideband cooling technique~\cite{Monroe1995} on the $|0\rangle\to|1\rangle$ transition with a global addressing beam. 
The measured mean number of phonons in all cooled modes, which can be  retrieved from the ratio of red and blue secular sidebands~\cite{Leibfried2003}, after the procedure is below 0.1. 

\subsubsection{Single-qudit operations}

\begin{figure}
\center{\includegraphics[width=0.4\linewidth]{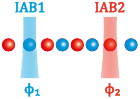}}
	\vskip-3mm
\caption{Due to difference in optical paths, relative phase between IAB1 and IAB2 $\phi_1-\phi_2$ slowly fluctuate between experimental shots. Therefore, to avoid nondeterminism of the relative phases of successive laser pulses applied to the ion, all particles are divided into two groups (shown as red and blue circles), each always controlled only with its own dedicated IAB (also shown as red and blue).}
\label{fig:ions_groups}
\end{figure}

The $\r{\phi}{0j}{\theta}$ operation is performed by applying a laser pulse resonant to the $\ket{0}\to\ket{j}$ transition with an individual addressing beam on a specific ion. 
The rotation axis $\phi$ is specified by the relative phase of the qudit and the phase of the laser beam and is controlled with the AOM. 
We note that right after the initialization the phase of the qudit is not defined and is fixed by the first quantum gate. An important issue is that two IABs when focused on the same ion have a slowly fluctuating phase difference due to slightly different optical paths. 
Thus, performing successive single-qudit gates on the same ion in the same quantum circuit with different beams would suffer from random phase errors and should be avoided. 
Therefore, all ions in the trap are divided into two groups, each always addressed with its own IAB, see Fig.~\ref{fig:ions_groups}.

A rotation angle $\theta=\Omega\tau$ is determined by the IAB-ion interaction Rabi frequency $\Omega$ and the pulse duration $\tau$. Empirically, the $\pi$-pulse duration of approximately 20~$\mu s$ has been found to be optimal to achieve the highest single-qudit gate fidelity.

The $\r{z}{j}{\theta}$ gate is implemented by the appropriate shift of the IAB phase during all successive real gates and always has a fidelity of 1.

\subsubsection{Two-qudit gate} \label{Sec:2Qg}

We perform the two-qudit MS gate by applying the bichromatic laser field to the target pair of particles with IABs. The field components are tuned close to the red and blue motional sidebands of radial modes along the $x$ trap axis. Disentangling of all motional modes from electronic degrees of freedom at the end of the gate is ensured by amplitude modulation of the laser field, similarly to Ref.~\cite{choi2014optimal} (so-called pulse shaping). 
The same modulation also makes the gate robust to experimental parameter fluctuations, such as secular frequency drifts. The shape consists of $2N+1$ segments ($N$ is the number of ions in the chain) of equal duration. The field amplitude during each segment is kept constant. Given the total duration of the gate and field detuning from the motional sidebands amplitudes of each segment can be found by solving the system of linear equations~\cite{choi2014optimal}, which close phase trajectories for each mode by the end of the gate and ensuring required gate parameter $\chi$. 
For each ion pair, we optimize gate parameters to maximize its speed and robustness to fluctuations within reasonable laser power levels. 
We experimentally calibrate gate parameters to perform gate for $\chi=\pi/4$, which is the most commonly used in our algorithms. 
However, we also interpolate the dependence of the $\chi$ on the used laser power and allow users to perform $XX$ gates with arbitrary $\chi$.

As it is mentioned above, relative phases between two IABs can slowly fluctuate on the timescale longer than a typical experimental shot due to the difference in optical paths. Because of that, all ions are divided into two groups, each always controlled only by its own IAB, see Fig.~\ref{fig:ions_groups}. 
While entangling two ions from different groups is straightforward, performing the two-qudit gate on a pair of ions from the same group implies illuminating one of them with an "incorrect" IAB (for example, a "red" ion with the "blue" IAB in the picture). 
This leads to the gate 
\begin{equation}
\exp\left(-\imath\frac{\chi}{2}(\sigma^{01}_x\otimes\mathbb{I}+\mathbb{I}\otimes\sigma^{01}_\phi)^2\right), 
\end{equation}
instead of 
\begin{equation}
\exp\left(-\imath\frac{\chi}{2}(\sigma^{01}_x\otimes\mathbb{I}+\mathbb{I}\otimes\sigma^{01}_x)^2\right),
\end{equation}
where $\phi$ fluctuates from shot to shot. 
In this case, we follow Ref.~\cite{Foss-Feig2021} and surround the entangling gate with single-qudit operations with the same IABs to convert it to $ZZ$ gate, which is insensitive to the relative qudit-laser phase.
As it is demonstrated in Ref.~\cite{nikolaeva2023universal}, with such a two-qudit operation can be obtained two-qubit operation between qubits embedded into different ququarts. 
This proves the universality of our gate set.

\subsubsection{Qudits readout}

\begin{figure}
\center{\includegraphics[width=0.5\linewidth]{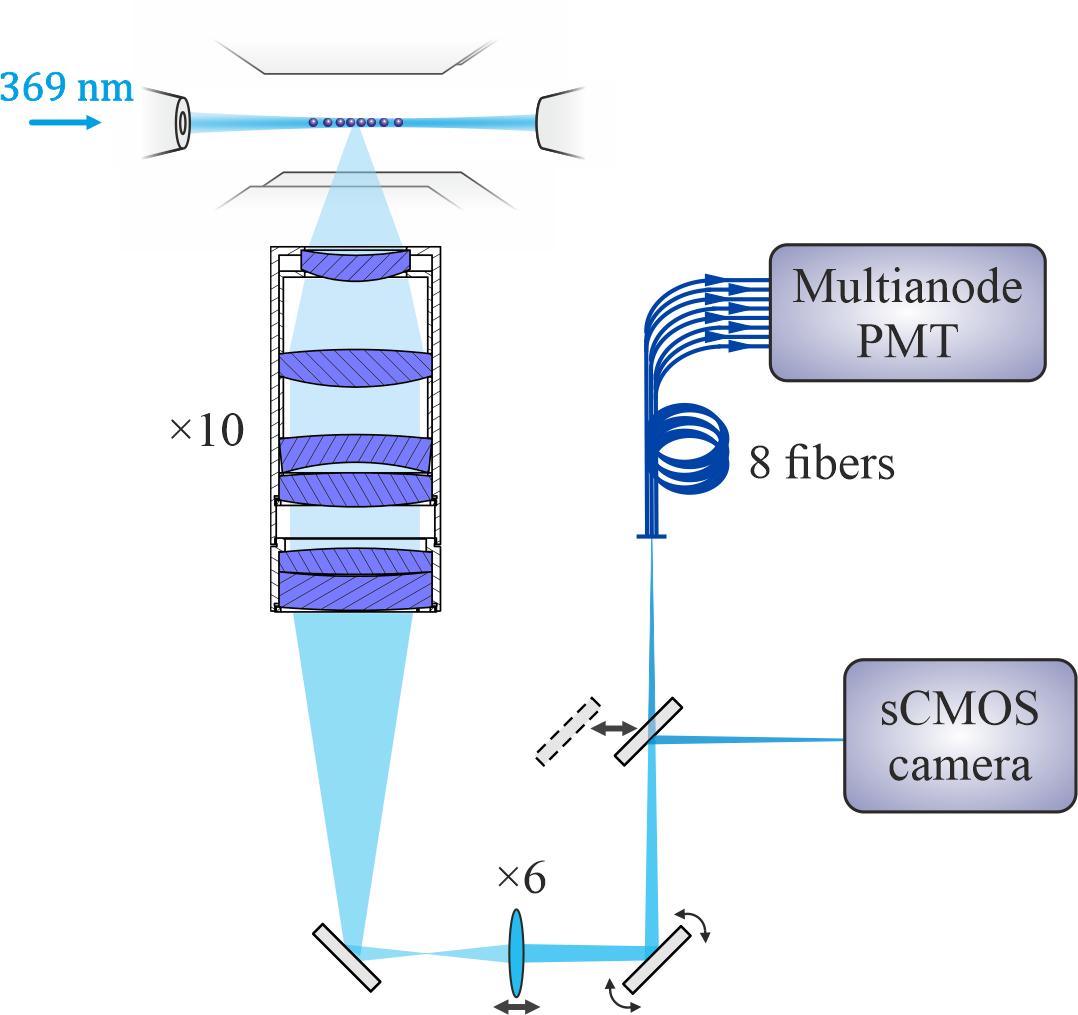}}
	\vskip-3mm
\caption{Readout optical scheme. Ions fluorescence is collected with a high-NA objective and focused onto either a sCMOS camera or an array of multimode fibers, coupled to a multichannel PMT. Detectors can be switched with a motorized mirror.}
\label{fig:readout}
\end{figure}

The readout of the qudits' states is performed with an electron shelving method~\cite{Leibfried2003} modified for multilevel systems. The first step of the readout procedure is the same as in the case of an optical qubit in $^{171}$Yb$^+$~\cite{ZalivakoSemerikov20212}. 
It is performed by turning on the cooling beam at 369~nm along the trap axis phase-modulated at 14.7~GHz and a repumping beam at 935~nm without phase-modulation for 1~ms. 
The repumping beam's intensity is kept below saturation to avoid non-resonant quenching of the excited qudits states. An objective with $NA=0.48$ collects ions fluorescence at 369~nm and creates a 10-fold magnified image of the ion chain, see Fig.~\ref{fig:readout}. 
This image is then further 6-fold magnified with a single lens and is projected onto an sCMOS camera or an array of multimode fibers depending on the position of a motorized translational stage with a mirror on it. 
In the latter case, each ion is coupled with its own fiber. 
The distances between fibers in the array correspond to the distances between ions in the chain. The other end of the fiber array is coupled to the multichannel photomultiplier tube (PMT). Current pulses from the PMTs are then multiplied and counted with the field programmable gate arrays (FPGA) that are capable of time-resolved pulses registering. 

If the ion is in the $|0\rangle$ state, we register a strong fluorescence from it, saying it is in the ``bright'' state. In all other cases (states $|1\rangle$, $|2\rangle$ or $|3\rangle$) the fluorescence will be absent (the ``dark'' state). 
We distinguish the ``bright'' and the ``dark'' states by comparing the number of registered photons from each ion with a pre-calibrated threshold value. We also note, that during this process, all population from the $|0\rangle$ state is pumped to the $^2S_{1/2}(F=1)$. 

To distinguish states $|1\rangle$, $|2\rangle$, and $|3\rangle$, we follow the protocols similar to described in Refs.~\cite{Ringbauer2021, Senko2020}. 
After the first readout stage, we apply to all ions a single-qudit operation $R_x^{01}(\pi)$, transferring all population from the state $|1\rangle$ to the $|0\rangle$ and repeat the state-dependent fluorescence detection. 
This time the ion will scatter photons if, in the end of the algorithm, it appears in states $|0\rangle$ or $|1\rangle$. 
The same actions are repeated one more time for the state $|2\rangle$. 
If the ion is bright for the first time during the measurement stage number $n$, the qudit is considered to be measured in the state $|n-1\rangle$. 
If after all three stages, 
it remains dark, it is considered to be in the state $|3\rangle$. 

\begin{figure}
\center{\includegraphics[width=0.5\linewidth]{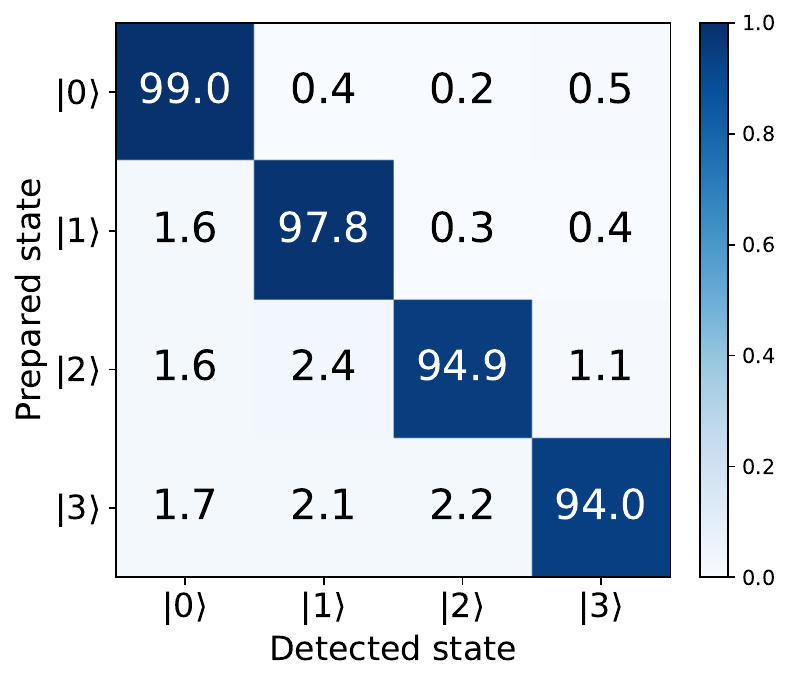}}
	\vskip-3mm
\caption{The SPAM confusion matrix averaged along all qudits in the register. Each row corresponds to a specific prepared state, while the columns correspond to the detected levels.}
\label{fig:spam}
\end{figure}

{In the Fig.~\ref{fig:spam} the SPAM fidelity of the described procedure is presented averaged along the ion chain. For this measurement each qudit was sequentially prepared to each of its states followed by the readout of the register. To prepare states $\ket{k}, k \ne 0$ the operation $\r{x}{0k}{\pi}$ were employed. The procedure was repeated for each qudit in the register and averaged. The main contribution to the readout error is a spontaneous decay of the excited qudit states during the measuring process (each readout stage takes 1~ms, while the upper states lifetime is 53~ms). This fidelity can be straightforwardly improved by improving the fluorescence detection efficiency and reducing the duration of the readout stage.}

This procedure can be readily extended for the readout of all possible 6 states of the $^{171}$Yb$^+$ qudit. 

\subsubsection{Experimental control}

The setup control system includes a set of digital input/output boards, direct digital synthesizers (DDSs), analog-to-digital converters (ADCs), digital-to-analog converters (DACs) and a master board based on the FPGA.
The latter ensures synchronization of all devices, feeds them command sequences required for each experimental shot, and registers input signals. 
The FPGA communicates with a computer via the high-speed Ethernet link where the main software is running. 
It also enables an operator to perform calibration of all components, runs quantum circuits, and is connected to the cloud platform. 
The platform allows users to build and remotely run their quantum circuits on the processor and analyze the results.

Multichromatic analog signals required for two-qudit operations are derived from several phase-coherent DDS channels, which are joined on a radiofrequency combiner. Amplitude pulse shaping is achieved by fast control over DDS signals from the FPGA.

\subsection{IQAE}\label{ref:appendix_iqae}
Here we provide a brief overview of the IQAE algorithm workflow. 
We transform fermionic operators in (8) in the main text into qubit operators, several transformations, such as  Jordan-Wigner~\cite{jw}, Parity~\cite{parity}, and Braviy-Kitaev~\cite{bk} transformations, can be used. 
Then the problem Hamiltonian, $H$, which consists of $N$-qubit terms $U_i$, can be presented as $H = \sum_{i=1}^r \beta_i U_i$, where $\beta_i \in \mathbb{C}$ are the Hamiltonian coefficients, and $r$ is the number of non-zero terms. 
The objective is to approximate the ground state and energy of $H$.
The algorithm is divided into three key stages:
\begin{enumerate}
    \item \textbf{Ansatz construction} Initially, we select a set of $L$ quantum states to form the ansatz $|\psi(\boldsymbol{\alpha})\rangle = \sum_{i=1}^{L}\alpha_{i}|\phi_{i}\rangle$. The parameters $\alpha_{i} \in \mathbb{C}$ are subsequently utilized in the optimization process. The states $|\phi_{i}\rangle$ constitute the Krylov subspace basis $\mathbb{C}\mathbb{S}^K = \{|\phi_i\rangle\} \cup \left\{ U_{i_1} |\phi_i\rangle \right\}_{i_1=1}^{r} \cup \dots \cup \left\{ U_{i_K} \dots U_{i_1} |\phi_i\rangle \right\}_{i_1=1,\dots,i_K=1}^{r}$,
    where $K$ defines the maximal order for the multiplication of Hamiltonian terms $U_{i}$.
    \item \textbf{Quantum computation} This step involves calculating the overlap matrices $D$ and $E$ of size $L \times L$ using quantum computer. The matrix elements are defined as  $D_{n, m}=\sum_{i=1}^{r}\beta_{i}\langle\phi_{n}|U_{i}|\phi_{m}\rangle$ and $E_{n,m} = \langle\phi_{n}|\phi_{m}\rangle$ . This step wraps up the quantum computer's role.
    \item \textbf{Classical processing} The obtained overlap matrices are processed using an optimization program on a classical computer: \textit{minimize} $\boldsymbol{\alpha}^{\dagger}D\boldsymbol{\alpha}$ \textit{subject to} $\boldsymbol{\alpha}^{\dagger}E\boldsymbol{\alpha}=1$.
    Alternatively, this can be formulated as a generalized eigenvalue problem, given by $D\boldsymbol{\alpha}=\lambda E\boldsymbol{\alpha}$.
    
\end{enumerate}

Let us show an example of how matrix $E$ can be obtained. Suppose we have the following Hamiltonian:
\begin{equation}
    H = IZ + ZI + ZZ + XX,
\end{equation} 
For simplicity, let's choose $|01\rangle$ as the initial state. 
Then, the first-order Krylov subspace $K=1$ basis vectors will be as follows (we multiply the Pauli strings from the Hamiltonian by the initial state vector once:
\begin{align}
    &\{II|01\rangle = |\phi_0\rangle, IZ|01\rangle = |\phi_1\rangle, \notag \\
    &ZI|01\rangle = |\phi_2\rangle, ZZ|01\rangle = |\phi_3 \rangle, XX|01\rangle = |\phi_4\rangle \},
\end{align}

For the second order (we multiply the Pauli strings from the Hamiltonian twice, select the unique ones, and multiply by the initial state vector):
\begin{equation}
    \{YX|01\rangle = |\phi_5\rangle, XY|01\rangle= |\phi_6\rangle, YY|01\rangle =|\phi_7\rangle\},
\end{equation}
In this case, the second order is the maximum order $K=2$; the vector space "closes", i.e., it is impossible to obtain new vectors by multiplying the existing vectors together.

If we limit ourselves to the first order, matrices 
$D$ and $E$ will have a dimension of $5 \times 5$ if we choose the second, the dimension of the matrices will be $8 \times 8$. 
The matrix $E$ for the first order looks like this:
\begin{equation}
E = \begin{bmatrix}
\langle \phi_0 | \phi_0 \rangle & \langle \phi_0 | \phi_1 \rangle & \langle \phi_0 | \phi_2 \rangle & \langle \phi_0 | \phi_3 \rangle & \langle \phi_0 | \phi_4 \rangle \\
\langle \phi_1 | \phi_0 \rangle & \langle \phi_1 | \phi_1 \rangle & \langle \phi_1 | \phi_2 \rangle & \langle \phi_1 | \phi_3 \rangle & \langle \phi_1 | \phi_4 \rangle \\
\langle \phi_2 | \phi_0 \rangle & \langle \phi_2 | \phi_1 \rangle & \langle \phi_2 | \phi_2 \rangle & \langle \phi_2 | \phi_3 \rangle & \langle \phi_2 | \phi_4 \rangle \\
\langle \phi_3 | \phi_0 \rangle & \langle \phi_3 | \phi_1 \rangle & \langle \phi_3 | \phi_2 \rangle & \langle \phi_3 | \phi_3 \rangle & \langle \phi_3 | \phi_4 \rangle \\
\langle \phi_4 | \phi_0 \rangle & \langle \phi_4 | \phi_1 \rangle & \langle \phi_4 | \phi_2 \rangle & \langle \phi_4 | \phi_3 \rangle & \langle \phi_4 | \phi_4 \rangle \notag
\end{bmatrix}  
\end{equation}

Suppose we want to calculate a certain matrix element 
$\langle \phi_4|\phi_1 \rangle$  of matrix $E$. 
We know that it can be reduced to a simpler form because our Hamiltonian is originally written as a sum of Pauli strings. Vectors $|\phi_4 \rangle = XX|01\rangle$, $|\phi_1 \rangle = IZ|01\rangle$, therefore,  $\langle \phi_4|\phi_1 \rangle = \langle 01 | XX * IZ|01\rangle$. 
Multiplying the corresponding position operators in the Pauli strings, we get $\langle \phi_4|\phi_1 \rangle = \langle 01 | -iXY |01\rangle = -i\langle 01 | XY |01\rangle$. 
The elements such as $\langle 01 | XY |01\rangle$ can be easily calculated using standard measurements in the computational basis.

\bibliography{bibliography.bib}